\documentclass[aps,prl,twocolumn,floatfix]{revtex4}
\pdfoutput=1
\usepackage{graphicx}
\usepackage{hyperref}

\usepackage{amssymb,amsfonts,amsmath}

\newcommand{\bra}[1]{\langle #1 |}
\newcommand{\ket}[1]{| #1 \rangle}
\newcommand{\figS}[1]{S#1}
\newcommand{\secS}[1]{S#1}

\begin{document} 

\title {Electron pairing by Coulomb repulsion in narrow band structures}
\author{Klaus M. Frahm and Dima L. Shepelyansky}

\affiliation{\mbox{Laboratoire de Physique Th\'eorique, IRSAMC, 
Universit\'e de Toulouse, CNRS, UPS, 31062 Toulouse, France}}

\date{February 16, 2020}


\begin{abstract}
{\bf Abstract:} 
We study analytically and numerically
dynamics and eigenstates of two electrons
with Coulomb repulsion on 
a tight-binding lattice in one and two dimensions.
The total energy and momentum of electrons
are conserved and we show that for a certain momentum range
the dynamics is exactly reduced 
to an evolution in an effective narrow energy band 
where the energy conservation forces the 
two electrons to propagate together 
through the whole system at moderate or
even weak repulsion strength.
We argue that such a mechanism of electron pair formation
by the repulsive Coulomb interaction is rather generic
and that it can be at the origin
of unconventional superconductivity in
twisted bilayer graphene.
\end{abstract}

\maketitle

{\it Introduction. -} The interactions of electrons 
in narrow band structures play an important role 
in various physical processes. Often the theoretical analysis
is done in the frame work of electrons on a lattice with 
short range on-site interactions as described in 
works of Gutzwiller and Hubbard 
\cite{gutzwiller1963,hubbard1963,gutzwiller1965}. 
Recently, the interest to narrow band structures
with strong Coulomb electron-electron interactions
has been inspired by the observation of unconventional
superconductivity in magic-angle
twisted bilayer graphene (MATBG) \cite{cao2018}.
Such structures are characterized by
a very high ratio of critical temperature of the superconducting transition 
to Fermi energy $T_c/E_F$ \cite{cao2018} and complex phase diagrams of 
superconducting and insulating phases \cite{ashoori2018nat,efetov2019nat}.
Experimental results for MATBG clearly show the importance of 
long range Coulomb
electron-electron interactions in these structures 
\cite{cao2018,ashoori2018nat,efetov2019nat,efetovscreen}.
From the theoretical side it has been shown that 
for small twisted angles the moir\'e pattern
leads to a formation of a superlattice with a unit cell 
containing more than 10000 atoms that significantly modifies 
the low-energy structure. Extensive numerical studies
by quantum-chemistry methods show the appearance of 
flat lowest-energy mini bands 
\cite{flatband1,flatband2,flatband3,flatband4}. 
These bands are rather narrow and thus the Coulomb interactions 
play an important role as it was already pointed out in early
theoretical studies \cite{macdonald}. The existence of narrow flat bands is
also confirmed in recent MATBG experiments \cite{berkeley}.

In this Letter we show that the narrow band structure of 
free electron spectrum leads to a number of 
unusual properties of their propagation in presence of 
strong, moderate or even weak Coulomb repulsion between electrons.
We discuss these properties on a model of two
electrons with Coulomb interaction propagating on 
a standard tight-binding lattice considered 
in \cite{gutzwiller1963,hubbard1963,gutzwiller1965}.
Our results show the appearance of pairing of
two electrons induced by a moderate Coulomb repulsion
with  ballistic pairs propagating over the whole system size.
Below we describe the physical properties of such pairs
of two interacting particles (TIP).

{\it Quantum tight-binding model of two electrons. -}
The quantum Hamiltonian of the model in dimensions $d=1$ or $2$ has the 
standard form 
\cite{gutzwiller1963,hubbard1963,gutzwiller1965}:
\begin{equation}
\label{eq_quant_Ham}
H=-\sum_{\langle j,l\rangle} \ket{j}\bra{l}+
\sum_j\frac{U}{1+r(j)} \ket{j}\bra{j}
\end{equation}
where $j=(x_1,x_2)$ ($j=(x_1,x_2,y_1,y_2)$) is a multi-index for $d=1$ 
($d=2$); each index variable takes values 
$x_1,x_2,y_1,y_2\in\{0,\ldots,N-1\}$ with $N$ being the linear system size
with periodic boundary conditions.
The first sum in (\ref{eq_quant_Ham}) describes the electron hopping
between nearby sites on a 1D (or 2D square) lattice
with a hopping amplitude taken as energy unit. 
The second sum in (\ref{eq_quant_Ham}) represents a (regularized) 
Coulomb type long-range 
interaction with amplitude $U$ and the distance $r(j)$ between 
two electrons. For 1D we have, due to the periodic boundary 
conditions, $r(j)=\Delta\bar x$ with 
$\Delta\bar x=\min(\Delta x,N-\Delta x)$ and relative coordinate 
$\Delta x=x_2-x_1$ which is taken modulo $N$ (i.e. $\Delta x=x_2-x_1+N$ 
for $x_2-x_1<0$). For 2D we have 
$r(j)=\sqrt{\Delta\bar x^2+\Delta\bar y^2}$. 
Furthermore, we consider symmetric (spatial) wave functions with respect to 
particle exchange assuming an antisymmetric spin-singlet state 
(similar results are obtained for antisymmetric wave functions).
In absence of interactions at $U=0$ the spectrum of free electrons 
has the standard band structure 
$E = -2\sum_{\mu=1,2;\,\alpha} \cos p_{\mu\alpha}$ 
with $\mu=1,2$ being the electron index and $\alpha=x$ in 1D 
($\alpha\in\{x,y\}$ in 2D) being the index for each spatial dimension.
With periodic boundary conditions each momentum $p_{\mu\alpha}$ 
is an integer multiple of $2\pi/N$.

{\it Classical dynamics of electron pairs. -} The corresponding 
classical dynamics in 2D is described by the Hamiltonian:
\begin{equation}
\label{eq_clas_Ham}
H = -2 \sum_{\mu=1,2;\,\alpha\in\{x,y\}} \cos p_{\mu\alpha} + 
U_C(x_1,x_2,y_1,y_2)
\end{equation}
with $U_C= U/[1+\sqrt{(x_1-x_2)^2+(y_1-y_2)^2}]$ and 
conjugated variables of momentum $p_{\mu x}, p_{\mu y}$ and coordinates
$x_\mu, y_\mu$ (in 1D we have in (\ref{eq_clas_Ham}) only $p_{\mu x}$ and 
$x_\mu$).
In 1D there are two integrals of motion being the total energy 
$E=H$ and total momentum $p_+=p_{1x}+p_{2x}$ leading to 
integrable TIP dynamics. In 2D we have 3 integrals of motion 
$E, p_{+x}, p_{+y}$ for 4 degrees of freedom and therefore 
the dynamics of the two electrons is generally chaotic
as it is shown in Fig.~\figS1 of \cite{suppmat}, Sec.~\secS1.  

Writing $\cos(p_{1x})+\cos(p_{2x})=2\cos(p_{+x}/2)\cos[(p_{2x}-p_{1x})/2]$ 
(and similarly for $y$) we see that at given values of $p_{+x},p_{+y}$ 
the kinetic energy is bounded by 
$\Delta E=4\sum_\alpha |\cos(p_{+\alpha}/2)|$. 
Therefore for TIP states with $E>\Delta E$ the two electrons 
cannot separate and they propagate as one pair. 
In particular for $p_{+x}=p_{+y}=\pi+\delta$ (with $|\delta|\ll 1$) being 
close to $\pi$ {\it there are compact Coulomb electron pairs
even for very small interactions $U $} as soon as 
$\Delta E \approx 2d|\delta|< U \ll B_d$ with $B_d=8d+U$ being the maximal 
energy bandwidth in $d$ dimensions. 
The center of mass velocity of such pairs (in direction $\alpha\in\{x,y\}$) is 
$v_{+\alpha}=(v_{1\alpha} + v_{2\alpha})/2=$
$2\cos(\delta/2)\,\sin(p_{1\alpha}-\delta/2)\approx 2 \sin p_{1\alpha}$
and it may be close to a maximal velocity $v_{+\alpha}=2$.
Fig.~\figS{1} of \cite{suppmat} clearly confirms the pair formation 
at small $U$ values and the pair propagation through the whole system. 

\begin{figure}[h]
\begin{center}
\includegraphics[width=0.4\textwidth]{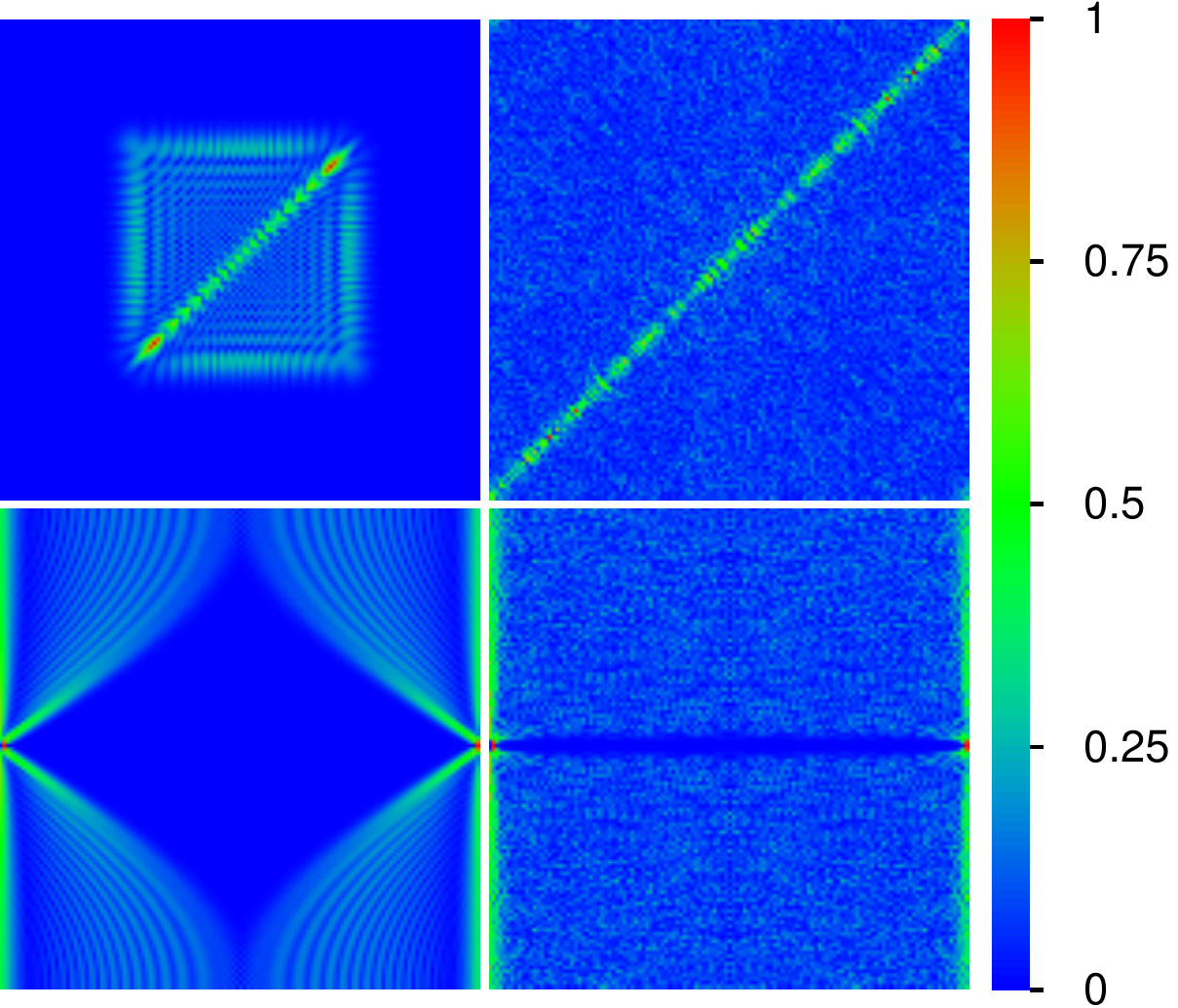}
\end{center}
\caption{\label{fig1} Color plot of wave function amplitudes obtained from 
the exact quantum time evolution in 1D at times 
$t=138\,\Delta t$ ($t=10^5 \Delta t$) in left (right) panels 
($\Delta t=1/B_1$ is an elementary time step of inverse band width $B_1=8+U$). 
Top panels show the TIP wave function amplitude 
$|\psi(x_1,x_2,t)|$ at $x_1$ and $x_2$ for both axis. 
Bottom panels show $|\bar\psi(p_+,\Delta x,t)|$ with $\Delta x = x_2-x_1$ 
(taken modulo N) being the relative coordinate (horizontal axis) and 
$0 \leq p_+ < 2\pi$ being the total momentum (vertical axis). 
The initial state at $t=0$ is 
a symmetrized state with one electron localized at 
$N/2$ and the other one at $N/2+1$, i.e. initial distance 
$\Delta \bar x=1$. Other parameters are $U=1$ and $N=128$;
color bar represents (here and in certain subsequent figures) 
the ratio of the shown quantity (here modulus of wave function amplitude)
to its maximal value.
Related videos are available at \cite{suppmat,ourwebpage}.
}
\end{figure}

{\it Quantum TIP eigenstates and dynamics in 1D and 2D. -} 
The factor $2\cos(p_+/2)$ for the kinetic energy (in 1D) 
at given value of $p_+$ can also be obtained from the 
quantum model. Using a suitable unitary transformation (see \cite{suppmat},  
Sec.~\secS2) one can transform the quantum Hamiltonian (\ref{eq_quant_Ham}) to 
a block diagonal form where the different blocks on the diagonal 
correspond, for each value of the conserved quantum number $p_+$, 
to an effective {\em one-particle} tight binding model (in $\Delta x$ space) 
with nearest neighbor hopping matrix element $-2\cos(p_+/2)$ and 
a diagonal potential given by $U/(1+\Delta\bar x)$ (see \cite{suppmat}, 
Sec.~\secS2 
for details, especially the boundary conditions and the generalization 
to the 2D case). Physically, this corresponds to a {\em single} particle 
in $\Delta x$ space with a kinetic energy rescaled by $2\cos(p_+/2)$ and 
moving in a given potential with maximal value $U$ for $\Delta x$ being close 
to $0$ or $N$. 

The eigenstates can be efficiently calculated for large system sizes since 
each block for a given value of $p_+$ can be diagonalized individually. 
Also the quantum time evolution (for $\hbar=1$) can be efficiently computed 
(up to $N=1024$) 
by transforming the initial state $\psi(x_1,x_2)$ into block representation 
$\bar\psi(p_+,\Delta x)$ and computing the time evolution exactly in 
each $p_+$ sector using the exact block eigenstates (details in 
\cite{suppmat}, Sec.~\secS2). 

A typical example of the TIP time evolution in 1D is shown in
top panels of Fig.~\ref{fig1} for initial electron positions (localized at 
$x_\mu\approx N/2$) at a distance $R=\Delta\bar x=1$ for $U =1$ and $N=128$. 
We see a wave front of free propagating separated electrons 
(square at small times in top left panel) and 
at the same time free propagation of the Coulomb electron pair along 
the diagonal $x_1=x_2$. At large times the density for separated electrons
is uniformly distributed  over the whole system while the density 
for Coulomb pairs is homogeneously distributed
over the whole diagonal keeping a relatively small pair size (top right panel).
Bottom panels show the block representation $|\bar\psi(p_+,\Delta x,t)|$ 
of the same states as in top panels. In this representation the initial state 
corresponds to two red vertical lines at $\Delta x=1$ and $\Delta x=N-1$ 
for perfect localization at these values and a uniform distribution 
in $p_+$ direction. With increasing time a part of the 
density stays quite well localized to the initial values with some modest 
increase of the initial width. The other fraction of the 
density propagates horizontally and becomes roughly uniform at sufficiently 
long times. The speed of the horizontal 
propagation is apparently proportional to $\cos(p_+/2)$ and for 
$p_+\approx \pi$ there is actually a small regime of strong, nearly perfect 
localization, where the weight of the propagating density is close to 0. 
In this case the kinetic energy can be considered as a small perturbation of 
the interaction due to the small ratio $2|\cos(p_+/2)|/U\ll 1$. However 
even for larger values of this ratio there is still a considerable fraction 
of the density that stays localized. 

\begin{figure}[h]
\begin{center}
\includegraphics[width=0.4\textwidth]{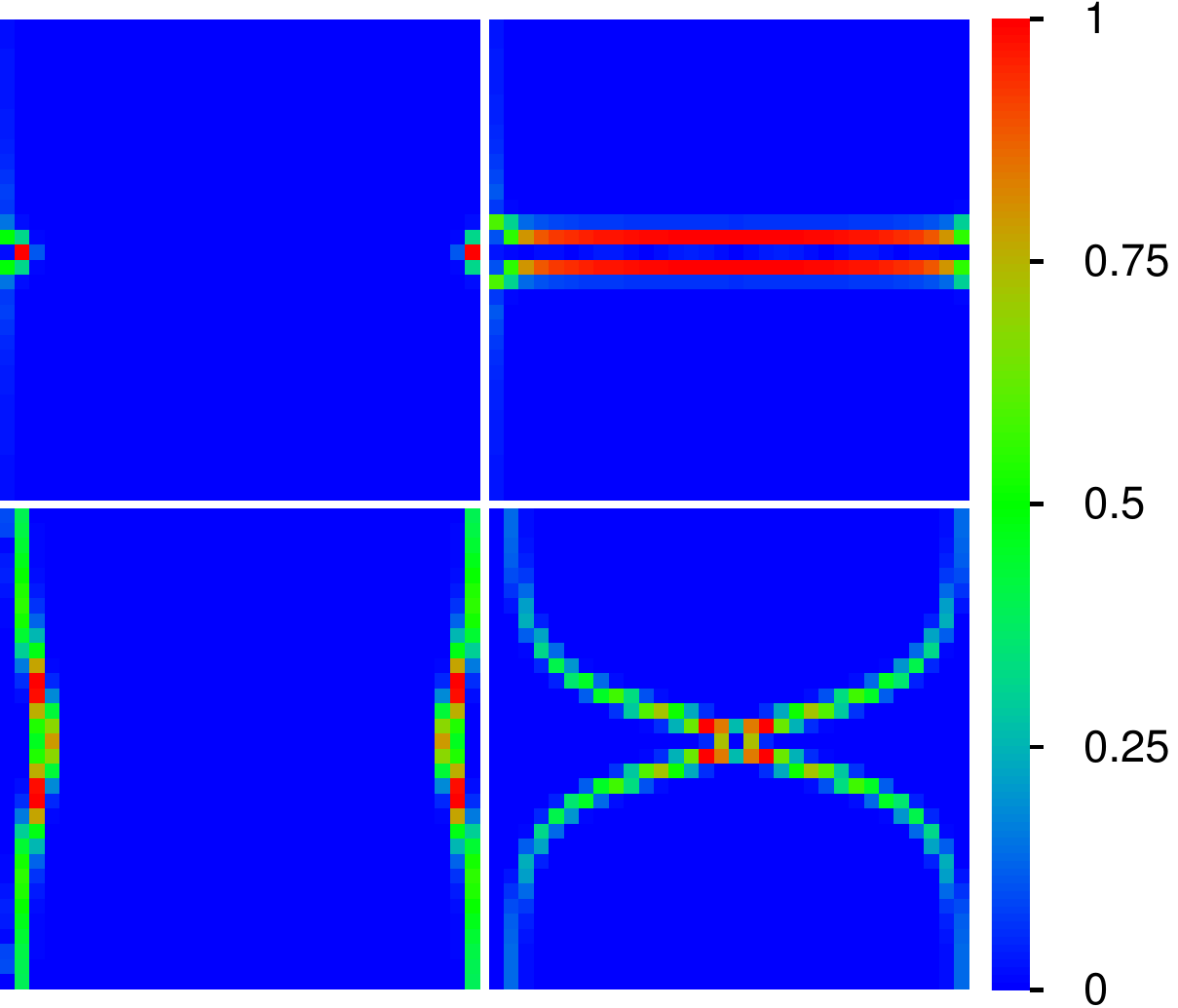}
\end{center}
\caption{\label{fig2} Examples of Husimi functions of symmetrized 
1D block eigenstates in relative coordinates 
$\Delta x\in[0,N[ $ ($x$-axis) and corresponding momentum 
$\Delta p\in[0,2\pi[$ ($y$-axis) in certain $p_+$ sectors 
for $N=1024$, $U=1$. 
Top (bottom) panels correspond to $p_+=341\pi/512\approx 2\pi/3$ 
($p_+=511\pi/512\approx \pi$).
Relative level numbers (ordered by increasing energies) inside each $p_+$ 
sector are 509 (top left), 474 (top right), 464 (bottom left), 374 (bottom 
right). 
}
\end{figure}

Fig.~\ref{fig2} shows certain Husimi functions of 1D block eigenstates 
obtained from the smoothing of the Wigner function on the scale of $\hbar$ 
(see e.g. \cite{husimi1,husimi2}) for two sectors $p_+\approx 2\pi/3$ and 
$p_+\approx \pi$. For larger energies there is a clear localization 
in $\Delta x$ at $\Delta x\approx 0$ and $\Delta x\approx N$ where the 
interaction potential is maximal and for medium/lower energies there is a 
near ballistic movement over the full $\Delta x$ range. For 
$p_+\approx 2\pi/3$ there is also localization in $\Delta p$ with two values 
close to $\pi$ in top right panel but the typical localized $\Delta p$ value 
may be different for other eigenstates (not shown) except for the few 
eigenstates with top energies which are localized in $\Delta x$. 
For $p_+\approx \pi$, where the effective kinetic energy is strongly reduced, 
there is a larger number of states with localization in $\Delta x$ for larger 
energies and for all cases $\Delta p$ appears to be rather delocalized. 
The states with localization in $\Delta x$ contribute mostly in the quantum 
time evolution shown in Fig.~\ref{fig1} due to the initial localized state at 
$\Delta x=1$ thus explaining the significant 
probability for both particles staying together. Further results of the 
1D quantum case are given in Figs.~\figS{2}-\figS{4} of 
\cite{suppmat}, Sec.~\secS3. 

\begin{figure}[h]
\begin{center}
\includegraphics[width=0.4\textwidth]{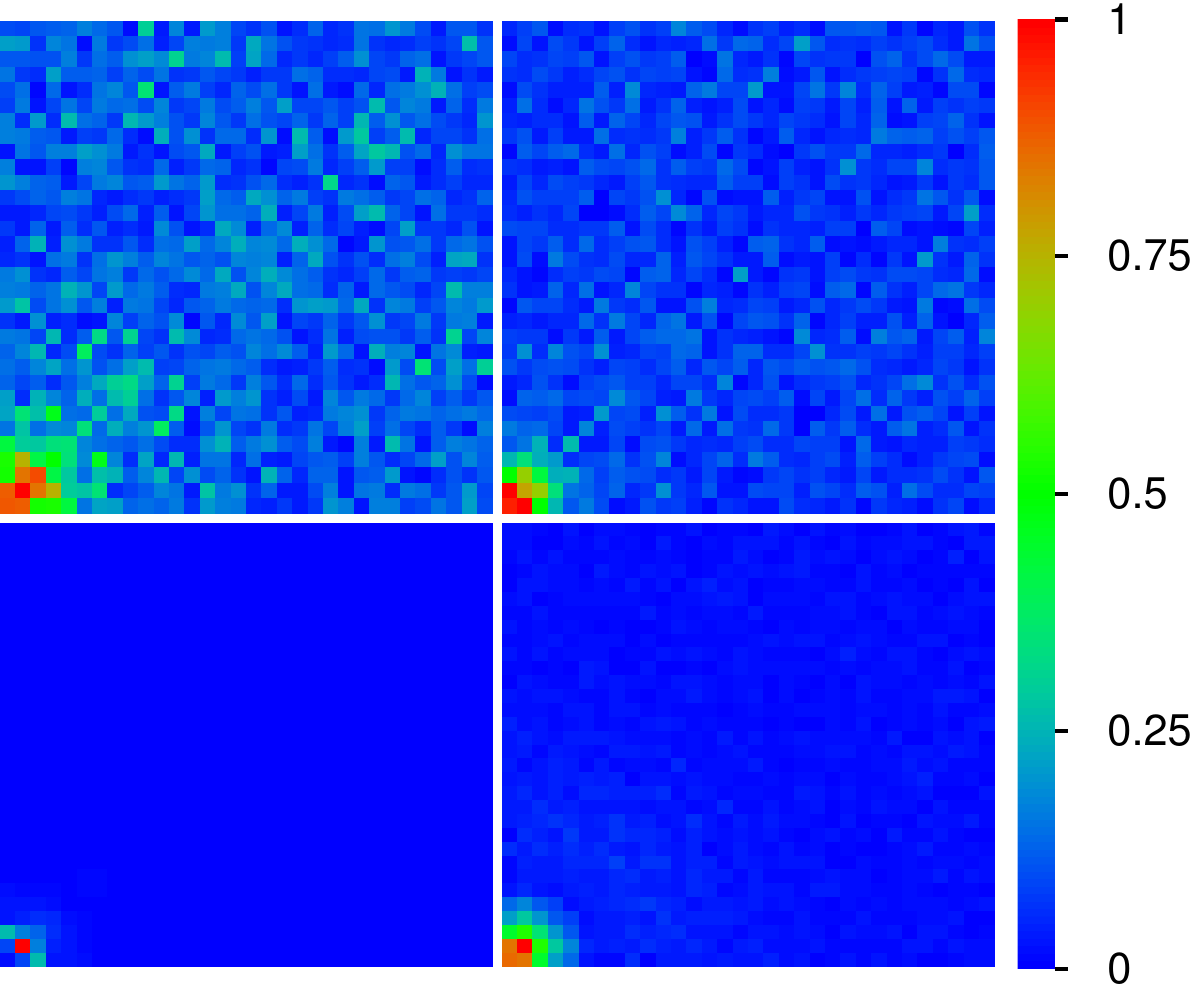}
\end{center}
\caption{\label{fig3} Color plot of wave function amplitude 
$|\bar\psi(p_{+x},p_{+y},\Delta x,\Delta y)|$ in block representation 
obtained from the 2D quantum time evolution at iteration time 
$t=10^5 \Delta t$ in $\Delta x$-$\Delta y$ plane 
for certain $p_{+x}$, $p_{+y}$ sectors and $U=2$ ($\Delta t=1/B_2=1/(16+U)$). 
The initial block state at $t=0$ is localized at $\Delta x=\Delta y=1$. 
All panels show a zoomed region $0\le \Delta x,\Delta y<32$. 
The values of  $p_+=p_{+x}=p_{+y}$ and $N$ are $p_+=0$, $N=128$ (top left), 
$p_+=21\pi/32\approx 2\pi/3$, $N=128$ (top right), 
$p_+=63\pi/64\approx \pi$, $N=128$ (bottom left), 
$k=85\pi/128\approx 2\pi/3$, $N=512$ (bottom right).
Related videos are available at \cite{suppmat,ourwebpage}.
}
\end{figure}

In order to access larger system sizes up to $N=512$ we compute for the 
2D case the exact quantum time evolution only in individual sectors with 
$p_{+x}=p_{+y}$ (and exploiting certain symmetries \cite{suppmat}).
The initial state is localized at $\Delta\bar x=\Delta\bar y=1$ but also 
at the chosen value of $p_{+x}=p_{+y}$. This corresponds 
to a wave in the center of mass direction with momenta $p_{+x},p_{+y}$ 
which is however perfectly localized in the relative coordinate. 

Fig.~\ref{fig3} clearly shows that there is a considerable probability 
that both electrons stay together even at very large times. There is however, 
for $p_{+x,y}=0$ or $p_{+x,y}\approx 2\pi/3$, also a certain complementary 
density for free electron propagation which becomes less visible for 
larger $N=512$. For $p_{+x,y}\approx \pi$ there is actually near 
perfect localization even at $N=128$. 
This confirms the formation of Coulomb electron pairs in the 2D quantum case 
and as in 1D also the particularly strong localization 
in relative coordinates for sectors with $p_{+x,y}\approx \pi$

\begin{figure}[h]
\begin{center}
\includegraphics[width=0.4\textwidth]{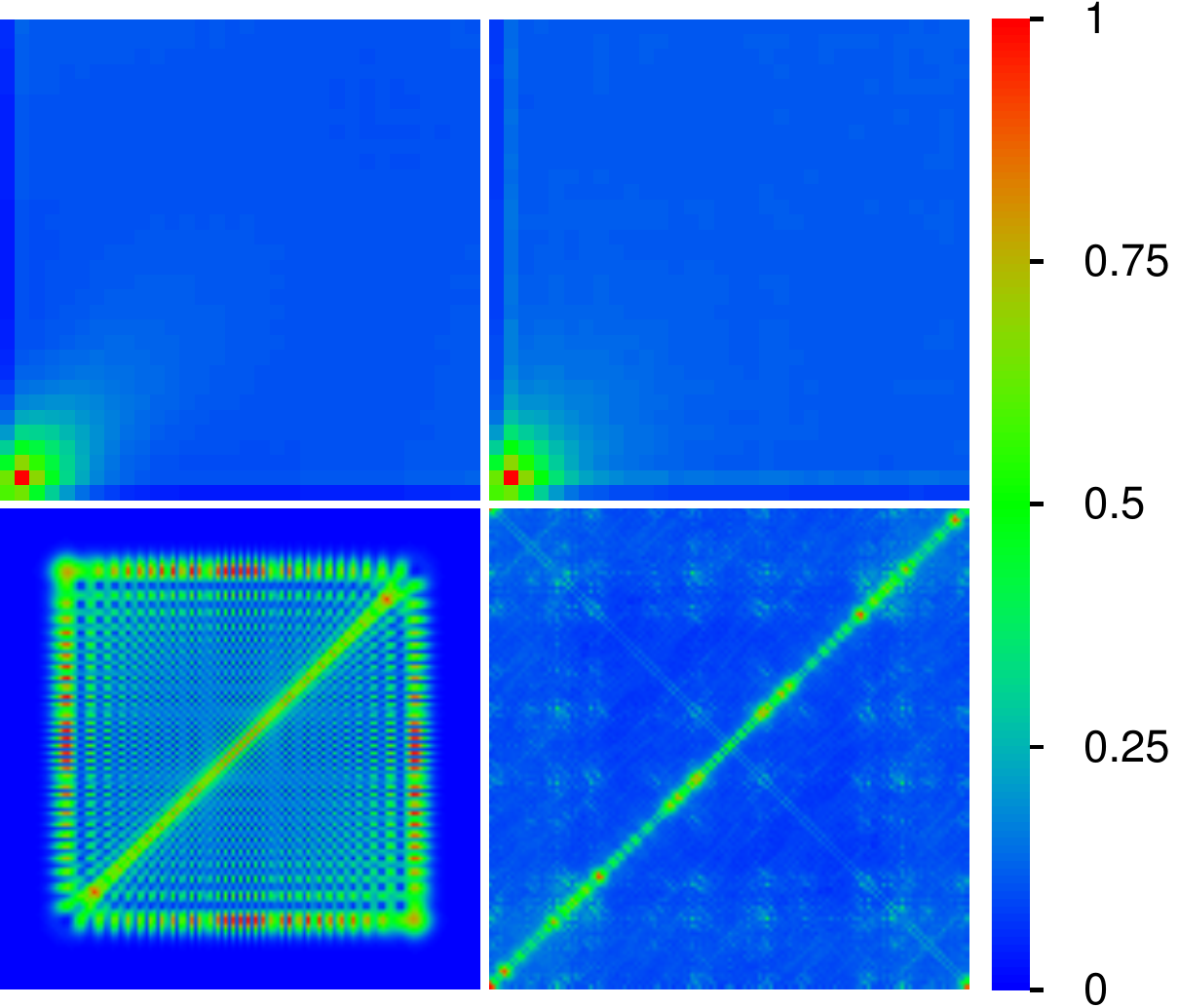}
\end{center}
\caption{\label{fig4} 2D Wave function densities obtained 
from the time evolution shown at times
$t=445\,\Delta t$ ($t=10^4 \Delta t$) in left (right) panels 
for initial electron positions at $\approx (N/2,N/2)$
and $N=128$, $U=2$ ($\Delta t =1/B_2=1/(16+U)$ is the Trotter integration 
time step). 
Top panels show the zoomed density for $0\le \Delta x,\Delta y<32$ 
in the $\Delta x$-$\Delta y$ plane of relative coordinates obtained from 
a sum over $x_1$ and $y_1$. Bottom panels show 
the density in $x_1$-$x_2$ plane obtained from a sum over $y_1$ and $y_2$. 
The corresponding value of probability near diagonal
$w_{10}$ is $w_{10}=0.106$ ($w_{10}=0.133$) for 
left (right) panels (see text). 
Related videos are available at \cite{suppmat,ourwebpage}.
}
\end{figure}

We also computed the full space 2D quantum time evolution using the Trotter 
formula approximation (see e.g. \cite{trotter} for computational details) 
with a Trotter time step $\Delta t=1/B_2=1/(16+U)=1/18$ for $U=2$ 
and for $N=128$. Here we choose an initial condition with both 
particles localized at $\approx (N/2,N/2)$ such hat 
$\Delta\bar x=\Delta\bar y=1$. 
The results of Fig.~\ref{fig4} clearly show that there is a 
significant probability of electron pair formation
with a propagation of pairs over the whole system 
(high probability of $\Delta x, \Delta y$ at values close to zero 
in top panels and near the diagonal $x_1 \approx x_2$ in bottom panels).
At the same time there is also a certain probability to have
practically independent electrons with ballistic propagation
through the lattice at moderate times (bottom left panel)
and an approximate homogeneous distribution over the whole system
at large times (bottom right panel).

In order to characterize the pair formation probability we 
compute the quantum probability $w_{10}(t)$ to find both 
electrons at a finite distance from each other
$\Delta\bar x\le \Delta R=10$ and $\Delta\bar y\le \Delta R$. 
At large times we find that the pair formation probability is 
$w_{10}=0.133$ while the probability to have independent electrons is 
$1-w_{10}$.

\begin{figure}[h]
\begin{center}
\includegraphics[width=0.4\textwidth]{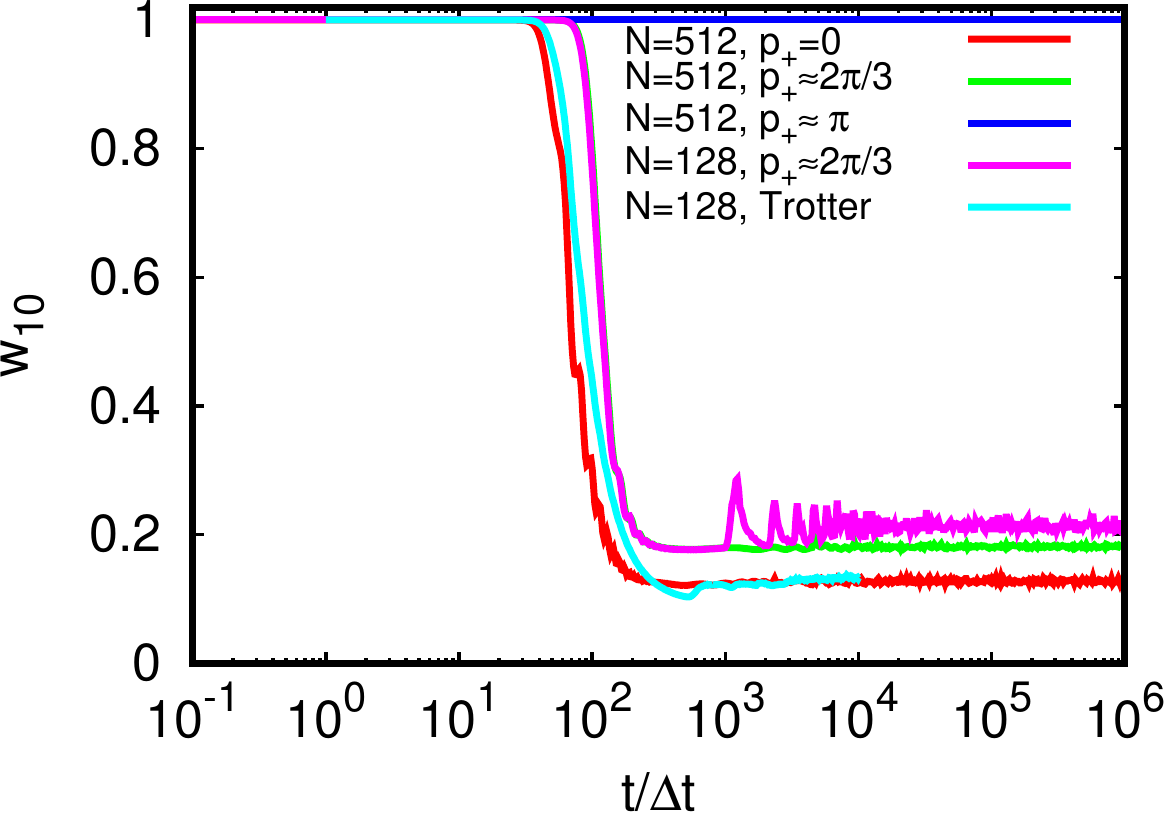}
\end{center}
\caption{\label{fig5} Time dependence of the pair formation probability 
$w_{10}$ for $U=2$ and different cases of either the exact 2D time evolution 
in certain sectors of conserved momenta $p_+=p_{+x}=p_{+y}$ 
or the full space 2D time evolution using the Trotter formula approximation 
(initial states as in Figs.~\ref{fig3},\ref{fig4}). The precise 
values of $p_+$ are~: $p_+=0$, $p_+=85\pi/128\approx 2\pi/3$, 
$p_+=255\pi/256\approx\pi$ all three for $N=512$ and 
$p_+=21\pi/32\approx 2\pi/3$ for $N=128$. 
The top blue line at $w_{10}=1$ is accurate with a numerical error 
below $10^{-14}$ for all time values.}
\end{figure}

The time dependence of the pair formation probability $w_{10}$ for both 
types of 2D quantum time evolution is shown in Fig.~\ref{fig5} for 
$U=2$ and $N\in\{128,512\}$. 
For $p_+ \approx \pi$ we have $w_{10}=1$ with numerical accuracy 
while for $p_+ \approx 2\pi/3$ and $p_+ =0$
this probability decreases with time but saturates at rather high values 
$w_{10} \approx 0.2$ and $0.13$ respectively.
The saturation value $w_{10} \approx 0.133$ for the full space Trotter 
formula approximation is very close to the case $p_+ =0$.

\begin{figure}[h]
\begin{center}
\includegraphics[width=0.4\textwidth]{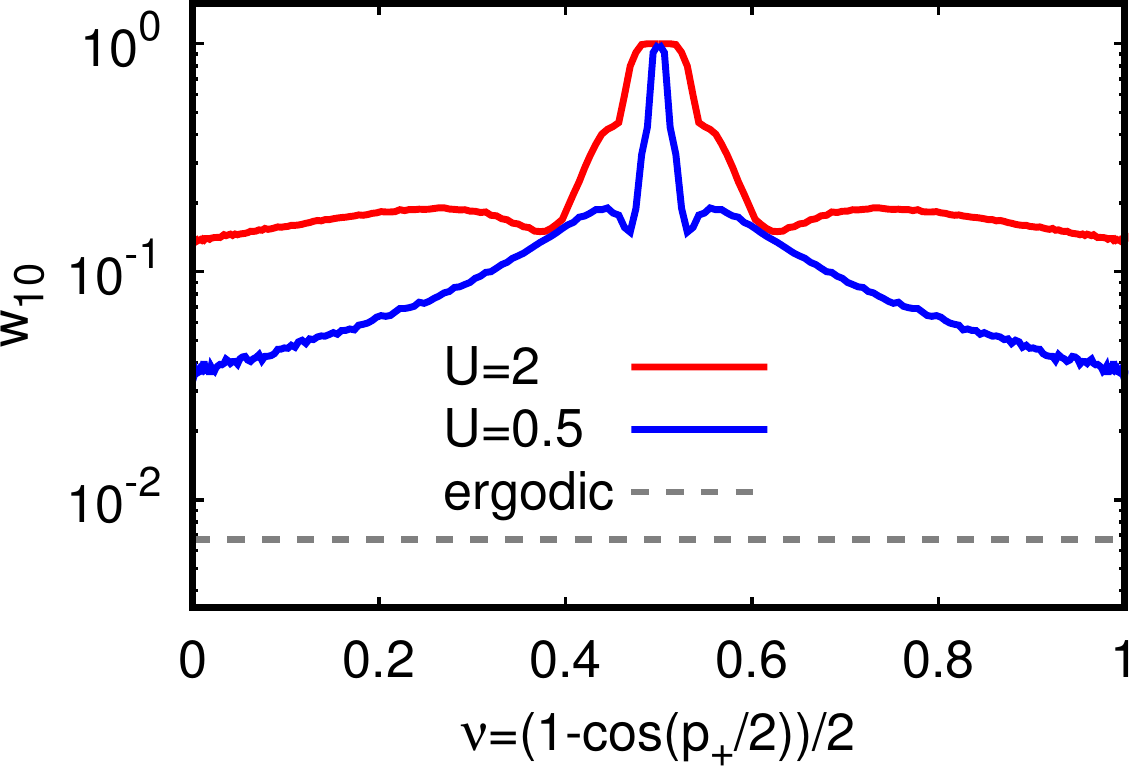}
\end{center}
\caption{\label{fig6} Dependence of pair formation probability 
$w_{10}$ on the filling factor $\nu$ for $U=2$ (red curve), $0.5$ (blue curve)
for $N=256$ in 2D. The dashed line shows the ergodic probability 
$w_{\rm erg.}$ assuming a uniform wave function. 
}
\end{figure}

The initial state with the pair momentum $p_+$
approximately corresponds to the electron Fermi energy
$E_F \approx - 4 \cos(p_+/2)$ assuming relatively 
moderate or weak interaction $U \ll B_d$. Thus the variation
of $p_+$ in the range $0 \leq p_+ \le 2\pi$
corresponds to the variation of the filling factor 
$\nu$ in the range $0 \leq \nu \leq 1$
with $\nu = (1-\cos(p_+/2))/2$. 
The dependence  $w_{10}(\nu)$ on $\nu$ is shown in Fig.~\ref{fig6} 
for $U=0.5,\,2$ and $N=256$ (data for $N=512$ is very close).
We see that for small $\nu \ll 1/2$ the 
electron pair formation probability is relatively small
but for $1/3 \leq \nu \leq 1/2$ it takes for $U=2$ ($U=0.5$) 
values from $w_{10} =0.17$ ($w_{10} =0.11$) to $w_{10}=1$. 
The values of $w_{10}$ are significantly above the ergodic 
probability $w_{\rm erg.}=(21/N)^2\approx 0.0067$
for $N=256$ assuming a uniform distribution.

More results for the quantum time evolution and eigenstates
properties in 2D are given in \cite{suppmat} 
(see Figs.~\figS{4}-\figS{10} of Sec.~\secS4).
In particular, there is a clear scaling dependence of $w_{10}$ (and other 
related characteristics) on the ratio $2\cos(p_+/2)/U$ 
(Figs.~\figS{9}, \figS{10} of \cite{suppmat})

{\it Discussion. -} The presented analytical and numerical
analysis definitely shows that a specific energy dispersion law
of free electrons in narrow bands leads to a new possibility
of pair formation induced by Coulomb repulsion between electrons.
The pairing of electrons already exists for a relatively moderate,
or even small, repulsion strength $U$. Of course, this analysis is 
performed in the frame work of only two interacting 
electrons. However, we conjecture that these
Coulomb electron pairs will still exist even at finite electron density $\nu$.
Indeed, a propagating pair can feel  interactions from other electrons
as a certain mean field potential which should not 
destroy pairs with a relative strong coupling.
In presence of an external space inhomogeneous potential
the independent electrons 
can be localized by the potential while the pairs,
being protected by their effective coupling energy
$\Delta_c \sim U/\Delta R$,
can remain insensitive to the potential
propagating through the whole system
(we assume that $\Delta R$ is  of the order of several lattice units
as for $w_{10}$ computations).
Thus such pairs can form a condensate leading to the 
emergence of a superconducting state.
We argue that the emergence of Coulomb electron pairs
in narrow band structures can be at the origin of
superconductivity in MATBG experiments.

{\it Acknowledgments. -}
This work was supported in 
part by the Programme Investissements
d'Avenir ANR-11-IDEX-0002-02, 
reference ANR-10-LABX-0037-NEXT (project THETRACOM).
This work was granted access to the HPC resources of 
CALMIP (Toulouse) under the allocation 2019-P0110.

\clearpage
\appendix 
\section{}
\label{appenda}
\setcounter{figure}{0} \renewcommand{\thefigure}{S\arabic{figure}} 
\setcounter{equation}{0} \renewcommand{\theequation}{S\arabic{equation}} 
\setcounter{page}{1}

\noindent{{\bf Supplementary Material for\\
\vskip 0.2cm
\noindent{Electron pairing by Coulomb repulsion in narrow band structures}}\\
\noindent by
K.~M.~Frahm and D.~L.~Shepelyansky.}

\vspace{0.5cm}
Here, we present additional material for the main part of the article.

\section{S1. Classical chaotic dynamics and propagation of two electrons in 2D}

\begin{figure}[h]
\begin{center}
\includegraphics[width=0.4\textwidth]{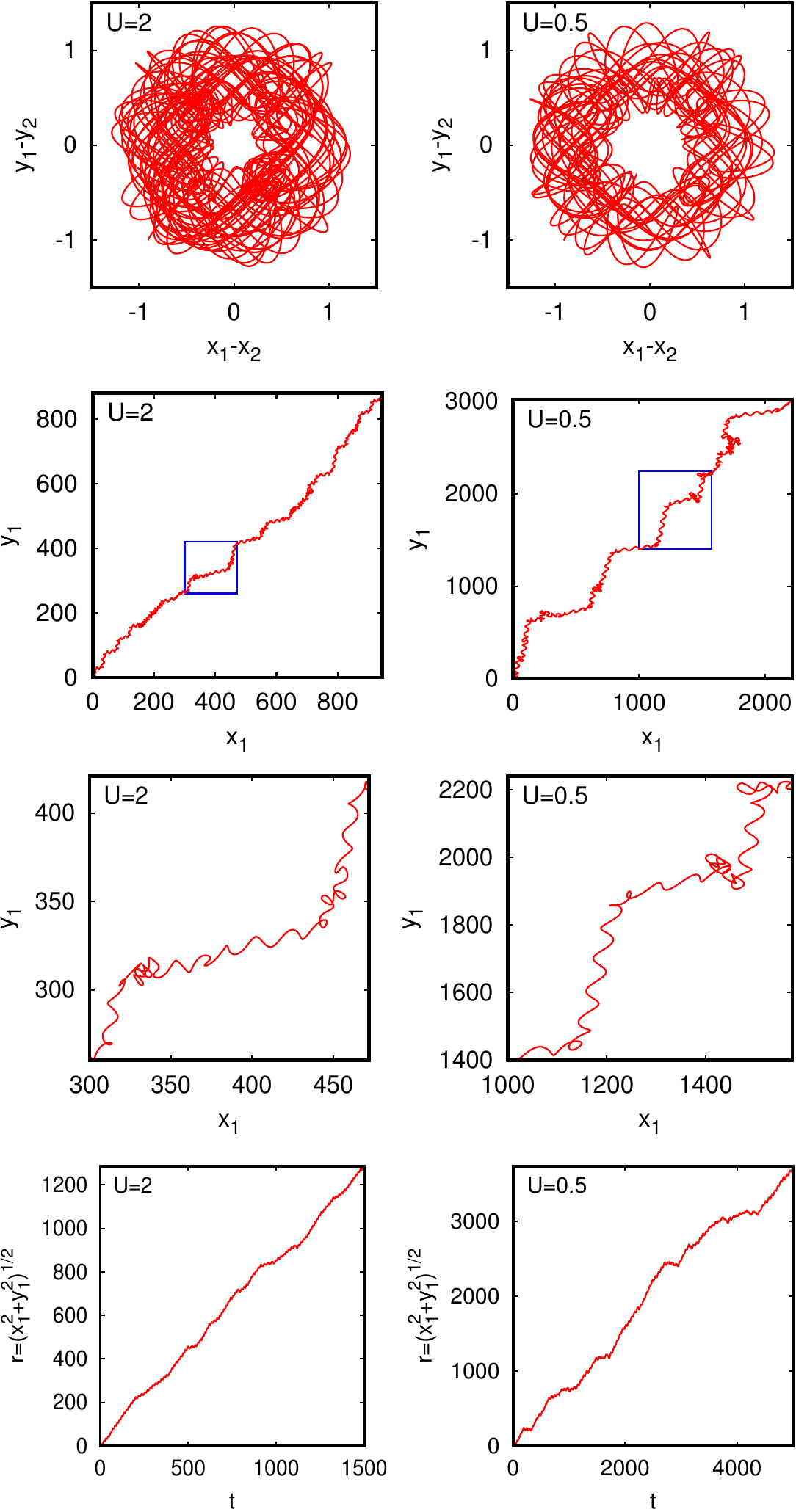}
\end{center}
\caption{\label{figS1} Classical dynamics for 2D case with a 
classical trajectory obtained by the standard Runge-Kutta 4th order method 
using a time step of $dt=0.005$. 
The initial condition corresponds to $|x_1-x_2|\approx |y_1-y_2|\approx 1$ 
and conserved total momenta being $p_{x1}+p_{x2}=p_{y1}+p_{y2}=\pi+\Delta p$ 
close to $\pi$. 
Panels of left (right) column correspond to $U=2$, $\Delta p=0.1$ and 
maximal time $t_{\rm max}=1500$ ($U=0.5$, $\Delta p=0.02$, 
$t_{\rm max}=5000$). Top panels show the dependence of $y_1-y_2$ on 
$x_1-x_2$; 2nd row of panels show the propagation of $y_1$ versus $x_1$ and 
3rd row of panels provide a zoomed view of the blue rectangles visible in 
2nd row panels revealing a loop like structures of the curves;
bottom panels show the time dependence of $r=\sqrt{x_1^2+y_1^2}$. 
}
\end{figure}

We show in Fig.~\ref{figS1} for $U=2$ and $U=0.5$ two 
typical trajectories obtained 
from the canonical equations with respect to the classical 2D Hamiltonian 
(\ref{eq_clas_Ham}) using initial conditions with 
$|x_1-x_2|\approx |y_1-y_2|\approx 1$ 
and conserved total momenta $p_{x1}+p_{x2}=p_{y1}+p_{y2}=\pi+\Delta p$ 
close to $\pi$. The projection of the trajectory on the 
$(x_1-x_2)$-$(y_1-y_2)$ 
plane clearly shows a bounded and chaotic dynamics
of two electrons. Obviously both electrons 
stay close together even though the interaction values are significantly 
smaller than the total energy width $B_2=16$ of two free electrons. 
The compact electron pair propagates quasi-ballistically through 
the whole system which can be seen from the global quasi-linear dependence 
of $y_1$ on $x_1$ and of $r=\sqrt{x_1^2+y_1^2}$ on $t$ even though 
there are some loop like structure on short length scales. 

\section{S2. Numerical methods for the computation of TIP eigenstates and 
quantum dynamics in 1D and 2D}

In this section we present some technical details about the properties of the 
quantum Hamiltonian 
(\ref{eq_quant_Ham}) and how to exploit its symmetries in order to 
efficiently compute eigenfunctions and solve exactly the quantum time 
evolution of this Hamiltonian. We remind that in (\ref{eq_quant_Ham}) 
the first sum for the kinetic energy is over nearest neighbors $j$ and $l$ 
such in both multi-indices exactly one of the four indices $x_1,x_2,y_1,y_2$ 
differs exactly by $\pm 1$ (or $\pm(N-1)$ if one index is 0 and the other 
$N-1$ according to the periodic boundary conditions) while the other three 
indices are equal.

We also remind the particular notation for the relative coordinate 
$\Delta x=(x_2-x_1+N)\mod N$ (and similarly for $\Delta y$), i.e. 
$\Delta x=x_2-x_1$ for $x_2\ge x_1$ and $\Delta x=x_2-x_1+N$ for $x_2<x_1$
such that $\Delta x\in\{0,\ldots N-1\}$. 
However, in order to correctly characterize a physical {\em distance} in the 
space of coordinates with periodic boundary conditions (e.g. for the 
interaction dependence on distance) it is the quantity $\Delta\bar x
=\min(\Delta x,N-\Delta x)$ (and similarly $\Delta\bar y$) which is 
relevant since a value of $\Delta x$ being close to $N$ 
corresponds in reality to a physical distance $N-\Delta x\ll \Delta x$.

\subsection{Discussion of 1D case}

In order to keep simpler notations we will use in the remainder this section 
the notation $k$ instead of $p_+$ for the total momentum (in 1D) and 
similarly $k_1,k_2$ for $p_1,p_2$ which is also more usual for momenta or 
wave numbers in the quantum case.

In the following we describe some technical details how to exploit the 
different symmetries for the 1D quantum case. If we denote by 
$\ket{k_1,\,k_2}$ a non-interacting eigenstate with $k_\mu=2\pi l_\mu/N$, 
$\mu=1,2$ and $l_\mu\in\{0,\ldots,N-1\}$ it is obvious that also 
in the presence of interaction the Hamiltonian (\ref{eq_quant_Ham}) only 
couples states such 
the total momentum $k=k_1+k_2$ is conserved leading to a block diagonal 
structure when expressing $H$ in the basis of non-interacting eigenstates. 

However, for numerical purposes this basis is not very convenient and we 
prefer to introduce a different basis of states given by:
\begin{equation}
\label{eq_block_basis}
\ket{k,\Delta x}=\frac{1}{\sqrt{N}}
\sum_{x_1=0}^{N-1} e^{ik(x_1+\Delta x/2)}\,\ket{x_1,\,x_1+\Delta x}
\end{equation}
where $\ket{x_1,\,x_1+\Delta x}$ corresponds to a basis vector 
in position space $\ket{x_1,\,x_2}$ with $x_2=x_1+\Delta x$ 
for $x_1,x_2,\Delta x\in\{0,\ldots,N-1\}$ (the sum $x_1+\Delta x$
is to be taken modulo $N$). 
Furthermore, $k=2\pi l_k/N$ corresponds to the total discrete momentum with 
$l_k\in\{0,\ldots,N-1\}$. The quantity 
$(x_1+\Delta x/2)\mod N$ corresponds to the center of mass (note that this 
quantity may be {\em different} from $(x_1+x_2)/2$ due to complications 
related to the periodic boundary conditions). In the following we call 
the states $\ket{k,\Delta x}$ block basis states. 
For a given state $\ket{\psi}$ we introduce in the usual way 
the wave functions $\psi(x_1,x_2)$ 
in position space and $\bar\psi(k,\Delta x)$ in combined $(k,\Delta x)$ space 
(also called {\em block representation}) by~: 
\begin{equation}
\label{eq_wavefunctions}
\ket{\psi}=\sum_{x_1,x_2}\psi(x_1,x_2)\,\ket{x_1,\,x_2}
=\sum_{k,\Delta x}\bar\psi(k,\Delta x)\,\ket{k,\,\Delta x}
\end{equation}
implying:
\begin{equation}
\label{eq_transform}
\bar\psi(k,\Delta x)=\frac{1}{\sqrt{N}}\sum_{x_1}
e^{-ik(x_1+\Delta x/2)}\,\psi(x_1,x_1+\Delta x)\ .
\end{equation}
One can easily work out that the Hamiltonian has also a block diagonal 
structure when expressed in terms of the block basis states, i.e. the 
block matrix elements are given by:
\begin{equation}
\label{eq_blockmatelem}
\bra{\tilde k,\Delta\tilde x}\,H\,\ket{k,\Delta x}=
\delta_{\tilde k,k}\,\bar h^{(k)}_{\Delta \tilde x,\Delta x}
\end{equation}
with symmetric $N\times N$ matrices 
$\bar h^{(k)}$, to be called {\em block Hamiltonians}, 
depending on $k$ and with non-vanishing 
matrix elements:
\begin{eqnarray}
\label{eq_block1}
\bar h^{(k)}_{\Delta x,\Delta x}&=&U(\Delta x)\ ,\\
\label{eq_block2}
\bar h^{(k)}_{\Delta x+1,\Delta x}&=&
\bar h^{(k)}_{\Delta x,\Delta x+1}
=-2\cos(k/2)\ ,\\
\label{eq_block3}
\bar h^{(k)}_{0,N-1}&=&
\bar h^{(k)}_{N-1,0}=-2 f_k \cos(k/2)\ ,
\end{eqnarray}
where (\ref{eq_block2}) applies to $0\le\Delta x<N-1$ and in (\ref{eq_block3})
$f_k=\exp(-kN/2)=(-1)^{l_k}$ with $l_k=kN/(2\pi)$ being the 
integer index associated to $k$. 
$U(\Delta x)=U/[1+\min(\Delta x,N-\Delta x)]$ is the interaction potential 
for the $1d$ case. 
The matrix $\bar h^{(k)}$ corresponds to a one-dimensional 
nearest neighbor tight binding Hamiltonian with {\em hopping matrix elements}: 
$-2\cos(k/2)$, {\em potential}: $U(\Delta x)$ and with {\em 
periodic (anti-periodic) boundary conditions} if $l_k$ is even (odd). 

This form is the quantum manifestation of the classical Hamiltonian when 
rewritten using the total momentum $p_+=p_1+p_2$, the relative coordinate 
$\Delta x=x_2-x_1$ with associated momentum $\Delta p=(p_2-p_1)/2$~:
\begin{equation}
\label{eq_hamclass1dP}
H_{\rm class}=-4\cos(p_+/2)\cos(\Delta p)+U(\Delta x)\ .
\end{equation}
The amplitude of the kinetic energy (hopping matrix element) is 
proportional to $2\cos(p_+/2)$ ($2\cos(k/2)$) and becomes very small 
for $p_+\approx\pi$ ($k\approx \pi$). At even $N$ there is actually one 
precise value $k=\pi$ where the hopping matrix element vanishes. 

Qualitatively, the eigenvectors of $\bar h^{(k)}$ at largest energies 
are expected to be localized at $\Delta x$ close to $0$ (or $N-1$) 
depending on the ratio $C=2|\cos(k/2)|/U$. 
If $C\ll 1$ there is even a perturbative regime with strong localization 
and for $k$ being sufficiently 
close to $\pi$ this scenario is already possible for modest interaction 
values such as $U=2$ or $U=0.5$ which are below the kinetic energy bandwidth 
$8$. Also for $C\sim 1$ the eigenfunction amplitudes at small $\Delta x$ 
are enhanced and a state initially localized at small $R=\Delta x$, e.g. 
$R=1$, will retain upon quantum time evolution an enhanced probability 
to stay close to small $\Delta x$ (or small $N-\Delta x$) 
values and propagate only partially to the remaining space.

These expectations are well confirmed by Fig.~\ref{fig2} showing 
for $U=2$, $N=1024$, the Husimi functions, defined 
in classical phase space of the relative coordinate $(\Delta x,\Delta p)$, 
(see \cite{husimi2} for precise definition and computational details using 
FFT) 
of certain eigenvectors $\varphi_l^{(k)}(\Delta x)$ of $\bar h^{(k)}$. The 
Husimi function for a given block eigenstate at energy $E_l^{(k)}$ is 
maximal on lines corresponding to the solution of 
$E_l^{(k)}\approx H_{\rm class}(\Delta x,\Delta p)$ with $H_{\rm class}$ is 
given by (\ref{eq_hamclass1dP}) at fixed value of conserved 
total momentum $p_+=k$. Depending on the energy $E_l^{(k)}$ 
there is either clear localization at 
$\Delta x$ close to $0$ or $N$ for larger energies and more or less free 
propagation for lower energies. If the ratio $C=2|\cos(p_+/2)|/U$ is 
small the range of localized states covers a larger energy interval and 
the remaining delocalized states feel stronger the shape of the interaction 
potential as can be seen in bottom-right panel of Fig.~\ref{fig2}. Note 
that in Fig.~\ref{fig2} we have chosen a quite large value of $N=1024$ 
in order to have a nice spatial resolution of the Husimi functions. 
We have actually also computed certain Husimi functions for even larger 
values $N=4096$ or $N=16384$, corresponding to smaller effective values 
of $\hbar$, which perfectly confirms the above observations but with 
a reduced width of the classical lines where the Husimi function is maximal. 

An important issue concerns the particle exchange symmetry implying that 
the eigenfunctions of the initial Hamiltonian are either symmetric (bosons) 
or anti-symmetric (fermions) 
with respect to particle exchange $x_2 \leftrightarrow x_1$. 
In the relative coordinate the symmetry should be visible by replacing 
$\Delta x$ with $(N-\Delta x)\mod N$. However, due to the discrete finite 
lattice with periodic boundary conditions the situation is more subtle. 
To see this we consider the particle exchange operator $P$ defined by 
$P\psi(x_1,x_2)=\psi(x_2,x_1)$. From (\ref{eq_transform}) we find that:
$P\bar\psi(k,0)=\bar\psi(k,0)$ and 
$P\bar\psi(k,\Delta x)=f_k\bar\psi(k,N-\Delta x)$ for $1\le \Delta x<N$ 
with the same $k$ dependent sign factor $f_k=\exp(-kN/2)=(-1)^{l_k}$ 
used in (\ref{eq_block3}). Therefore a symmetric (anti-symmetric) 
state $\varphi(\Delta x)$ (in $\Delta x$ space) 
with $P\varphi=s\,\varphi$, $s=+1$ ($s=-1$) 
satisfies $\varphi(0)=s\varphi(0)$ and 
$\varphi(\Delta x)=s\,f_k\varphi(N-\Delta x)$ 
for $1\le \Delta x<N$ implying $\varphi(0)=0$ for the anti-symmetric case 
and also $\varphi(N/2)=0$ for even $N$ and the anti-symmetric (symmetric) 
case for $f_k=1$ ($f_k=-1$). 

The block Hamiltonian (\ref{eq_block1}-\ref{eq_block3}) 
obviously commutes with $P$ and using a basis of symmetrized and 
anti-symmetrized states in $\Delta x$ space according to the above 
transformation rule one can transform $\bar h^{(k)}$ in a diagonal 
$2\times 2$ block structure for the two symmetric and anti-symmetric sectors. 
The dimension of the symmetric sector is 
$N_S=(N+1+f_k)/2$ for even $N$ and $N_S=(N+1)/2$ for odd $N$ and 
the anti-symmetric sector has the dimension $N_A=N-N_S$. 
The dependence of $N_S$ on the sign factor $f_k$ for even $N$ is due to the 
fact that the basis 
state $\ket{\Delta x}$ at $\Delta x=N/2$ either belongs to the symmetric 
sector for even $l_k$ (with $k=2\pi l_k/N$) or to the anti-symmetric sector 
for odd $l_k$ while for $\Delta x=0$ it always belongs to the 
symmetric sector. 
In all other cases for $0<\Delta x<N/2$ the (anti-)symmetrized basis states 
are $(\ket{\Delta x}+sf_k\ket{N-\Delta x})/\sqrt2$ with $s=+1$ ($s=-1$). 
The matrix elements of the (anti-)symmetrized block Hamiltonian are similar to 
(\ref{eq_block1}) and (\ref{eq_block2}) with some slight modifications: 
(i) the coupling matrix element (\ref{eq_block2}) implicating the 
states $\ket{0}$ and $\ket{N/2}$ (for even $N$) if they belong to the 
sector are multiplied with $\sqrt{2}$, (ii) there is no corner matrix element
corresponding to (\ref{eq_block3}) and (iii) for odd $N$ one 
has to add $-2sf_k\cos(k/2)$ to the diagonal matrix element (\ref{eq_block1}) 
for $\Delta x=(N-1)/2$. 

Our aim is to numerically compute the quantum time evolution of a state 
$\ket{\psi(t)}=e^{-iHt}\ket{\psi(0)}$ with initial state $\ket{\psi(0)}$ which 
we choose to be symmetric with respect to particle exchange. 
An exact way for this is to diagonalize $H$ (eventually in its 
symmetric sector) 
and expand $\ket{\psi(0)}$ in the basis of (symmetrized) eigenstates. 
Without any further optimization this requires 
${\cal O}(N^6)$ operations for the diagonalization and ${\cal O}(N^4)$ 
operations for each time value $t$ for which $\ket{\psi(t)}$ is computed 
(the limitation to the symmetrized sector reduces the numerical prefactors 
by 8 or 4). 

However, using the block Hamiltonians and the block diagonal structure 
of (\ref{eq_blockmatelem}) we can simplify the numerical diagonalization 
significantly by diagonalizing $N$ (or $\sim N/2$, see below) $N\times N$ 
matrices (or even $N/2\times N/2$ matrices in the symmetric sector) which can 
be done with $N{\cal O}(N^3)$ operations (or even $N{\cal O}(N^2)$ operations 
when exploiting the tridiagonal structure of the symmetrized block 
Hamiltonian of each sector). Once the eigenfunctions 
$\varphi_l^{(k)}(\Delta x)$ of $\bar h^{(k)}$ 
are known one can reconstruct the eigenfunctions $\phi_l^{(k)}(x_1,x_2)$ 
of the initial Hamiltonian $H$ by the inverse of the 
transformation (\ref{eq_transform}):
\begin{equation}
\label{eq_eig_reconstruc}
\phi_l^{(k)}(x_1,x_2)=\frac{1}{\sqrt{N}}\,e^{ik(x_1+\Delta x/2)}
\,\varphi^{(k)}(\Delta x)
\end{equation}
with $\Delta x=(x_2-x_1+N)\mod N$. Note that usually the inverse 
transformation from 
block to position representation for an arbitrary state would also require a 
sum over $k$. However, for eigenstates there is only one $k$ sector with 
non-vanishing values and the value of $k$ is simply fixed. 
From the numerical point of view the eigenstate (\ref{eq_eig_reconstruc}) 
is not convenient 
since it is complex (for $k\neq 0$ and $k\neq\pi$). Therefore we take 
the real and imaginary part which provides actually {\em two eigenstates} 
that are linear combinations of the eigenstates (\ref{eq_eig_reconstruc}) 
at $k$ and $2\pi-k$. This is related to the time reversal symmetry 
which provides the identity: $\bar h^{(2\pi -k)}=T\bar h^{(k)}T$ where 
$T$ is a diagonal matrix in $\Delta x$ space with non-vanishing elements 
$T_{\Delta x,\Delta x}=(-1)^{\Delta x}$. Due to this $\bar h^{(2\pi -k)}$ 
and $\bar h^{(k)}$ have the same eigenvalues and the operator $T$ provides the 
transformation of the eigenvectors from the latter to the former. Therefore 
it is only necessary to diagonalize $N/2$ of the $N$ block Hamiltonians 
numerically. For $k=0$ and $k=\pi$ (for even $N$) the eigenstates 
(\ref{eq_eig_reconstruc}) are already real (or purely imaginary) 
and for $k=\pi$ the 
block Hamiltonian is already diagonal since in this case the hopping matrix 
element (\ref{eq_block2}) vanishes. 

In this way, we can very efficiently construct a basis of real eigenstates 
of the Hamiltonians which provides a significant acceleration of the 
time evolution computation. However, for larger values of $N$ we still 
need a lot of storage for all eigenstates and also the transformations 
between the initial basis and the eigenstate basis are still quite expensive. 
Therefore, we apply a further optimization where we transform the initial 
vector $\ket{\psi(0)}$ in block representation using (\ref{eq_transform})
and then we apply the time evolution individually for each sector of $k$ which 
is highly efficient and furthermore it only requires to store the 
``small'' block eigenstates $\varphi_l^{(k)}(\Delta x)$ . The 
state $\ket{\psi(t)}$ is transformed back from block to position presentation 
(in addition both transformations can also be accelerated by FFT). 
A further advantage is 
that we can also analyze easily the state $\ket{\psi(t)}$ in block 
representation which is physically very interesting since it shows 
the (absence or presence of) propagation or localization 
individually for each $k$ sector 
corresponding to different hopping matrix elements ``$-2\cos(k/2)$'' 
(see bottom panels of Fig.~\ref{fig1}). Using this very highly efficient 
method we have been able to compute the {\em exact} full space 1D time 
evolution up to $N=1024$ corresponding to a Hilbert space dimension 
$\approx 10^6$. 
We have also verified that the three variants (and further sub-variants 
with respect to symmetry) of the method provide 
identical results up to numerical precision for 
sufficiently small values of $N$ where all methods are possible. 

\subsection{Discussion of 2D case}

The block representation (\ref{eq_transform}) can be generalized to the 
2D case providing wave functions $\bar\psi(k_x,k_y,\Delta x,\Delta y)$ 
depending on two total conserved momenta $k_x,k_y$ in $x$ or $y$-direction, 
$\Delta x$ and $\Delta y$. In this case the matrix elements in 
block basis states provide also a block diagonal structure:
\begin{eqnarray}
\label{eq_blockmatelem2D}
&&\bra{\tilde k_x,\tilde k_y,\Delta\tilde x,\Delta\tilde y}\,H\,
\ket{k_x,k_y,\Delta x,\Delta y}=\\
\nonumber
&&\qquad\qquad\quad\delta_{\tilde k_x,k_x}\delta_{\tilde k_y,k_y}
\,\bar h^{(k_x,k_y)}_{(\Delta \tilde x,\Delta\tilde y),(\Delta x,\Delta y)}
\end{eqnarray}
with symmetric $N^2\times N^2$ matrices $\bar h^{(k_x,k_y)}$ corresponding 
to a 2D-tight binding model in $(\Delta x,\Delta y)$ space 
with diagonal matrix elements given 
by $U(\Delta x,\Delta y)=U/(1+\sqrt{\Delta\bar x^2+\Delta\bar y^2})$ and 
nearest neighbor coupling matrix elements $-2\cos(k_x/2)$ ($-2\cos(k_y/2)$) 
in $x$ ($y$) direction. The boundary conditions are either periodic or 
anti-periodic in $x$ (or $y$) direction according to the parity of 
the integer number $Nk_x/(2\pi)$ ($Nk_y/(2\pi)$). 

The two 
discrete symmetries $\Delta x\leftrightarrow N-\Delta x$ (corresponding to 
$x_1\leftrightarrow x_2$) and $\Delta y\leftrightarrow N-\Delta y$ allow 
to simplify the diagonalization problem to matrices of size $\approx N^2/4$. 
(Note that the particle exchange symmetry corresponds to the simultaneous 
application of {\em both} of these symmetries.) 
If $k_x=k_y$ one can even exploit a third symmetry with respect to 
$\Delta x\leftrightarrow \Delta y$ allowing a further reduction of the 
matrix size to $\approx N^2/8\approx 3.3\times 10^4$ for $N=512$ which is 
still accessible to simple full numerical diagonalization. 
The additional symmetries with respect to the particle exchange 
symmetry are due to the particular simple form of the initial tight-binding 
model given as a simple sum of 1D tight-binding models for each direction 
(for the kinetic energy). 

The details with many different cases for the parity of both $k_x,k_y$ values, 
and also of $N$, together with several cases of symmetric or anti-symmetric 
sectors are quite complicated. For the time evolution of the 
2D case we limit ourselves to a single sector with $k=k_x=k_y$, even $N$ 
and also to the totally symmetric case (with respect to the three exchange 
symmetries mentioned above). In this case each (totally 
symmetrized) block Hamiltonian has a 
dimension $D_{k}=(N+1+f_{k})(N+3+f_{k})/8\approx N^2/8$ where 
$f_{k}=(-1)^{l_k}$ is the sign factor associated to $k$. As initial state 
we use a totally symmetrized state localized at 
$\Delta\bar x=\Delta\bar y =R$ where we 
choose typically $R=1$ but we have also performed certain computations 
for $R=3$ (see e.g. Fig.~\ref{figS8} below).
In the original full 4D space of TIP such a state corresponds to a plane 
wave in the center of mass direction and perfect localization in the relative 
coordinate. Therefore, the initial spreading along the diagonal 
(center of mass coordinate) is not visible with such a state 
(contrary to the full space 1D time evolution where an initial state 
localized in the center of mass was used as can be seen in top panels of 
Fig.~\ref{fig1}). However the more important 
spreading in the relative coordinate, which determines the pair formation 
probability, is of course clearly visible. 

In addition to the exact block 2D time evolution we also computed the full 
space time evolution using the Trotter formula approximation:
\begin{eqnarray}
\label{eq_trotter}
&&\exp[-i(H_{\rm kin}+H_{\rm pot})t]\approx\\
\nonumber
&&\qquad
[\exp(-iH_{\rm kin}\Delta t)\exp(-iH_{\rm pot}\Delta t)]^{t/\Delta t}
\end{eqnarray}
which is valid 
for a sufficiently small Trotter time step $\Delta t$ and time values $t$ 
being integer multiples of $\Delta t$. Here $H_{\rm kin}$ represents 
the kinetic energy part of the Hamiltonian (diagonal in Fourier space) 
and $H_{\rm pot}$ represents the part with potentials and interactions 
(diagonal in initial position space). Using a 4D-FFT to transform efficiently 
between initial position space and Fourier space one can compute the
approximate time evolution of an arbitrary given initial state. Further 
technical details of this method for the case of 2D TIP can be found 
in \cite{trotter}. 

As Trotter time step we choose the inverse 
bandwidth $\Delta t=1/B_2=1/(16+U)$ (which is below $\Delta t=0.1$ 
used in \cite{trotter}). Also for the exact time evolution (in 1D and 2D) 
we measure/present all time values in units of the inverse bandwidth 
$\Delta t=1/B_d$ which corresponds to the smallest time scale 
of the system. However, for the latter the ratio $t/\Delta t$ may 
be arbitrary and is not limited to integer values as for the 
Trotter formula approximation. 
Concerning the Trotter formula time evolution we choose a system size of 
$N=128$ (as in \cite{trotter}) and an initial state being roughly localized at 
$\approx N/2$ for each of the four coordinates with exact initial 
distance $\Delta\bar x=\Delta\bar y=R$ and totally symmetrized with respect 
to the above mentioned three discrete symmetries. 

The considered time range for all types of quantum time evolution 
computations is 
$\Delta t\times 10^{-1}\le t\le \Delta t\times 10^6$ 
($\Delta t\le t\le \Delta t\times 10^4$) for the 
exact 1D/2D (Trotter formula 2D) 
time evolution using an (approximate) logarithmic scale for the chosen 
time values. We provide here two example videos and 
at \cite{ourwebpage} further videos (see Sec.~S5 for details) 
for several of the densities shown in Figs.~\ref{fig1},\ref{fig3},\ref{fig4} 
and \ref{figS7}.

\section{S3. Additional data for the 1D quantum case}

\begin{figure}[h]
\begin{center}
\includegraphics[width=0.4\textwidth]{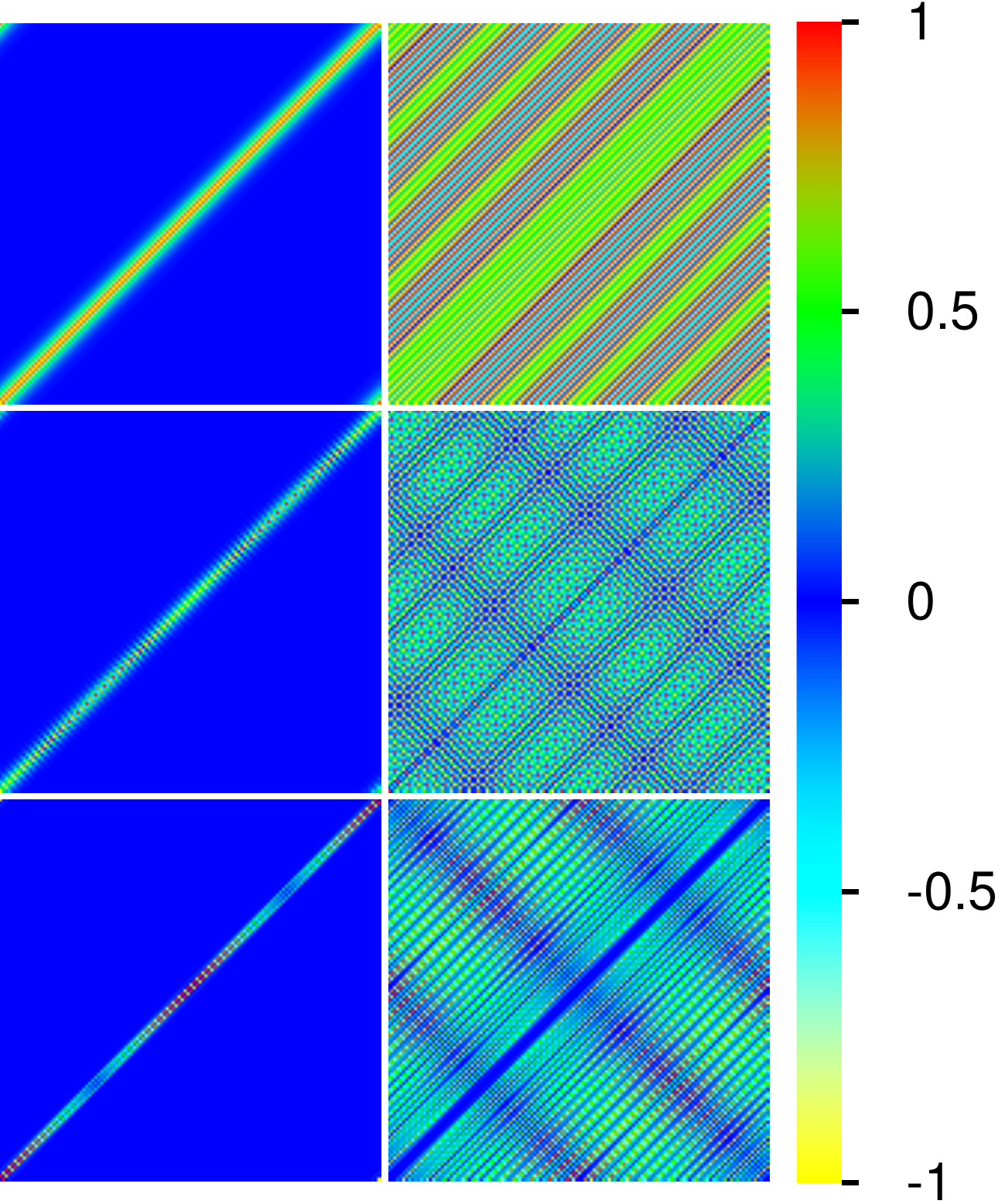}
\end{center}
\caption{\label{figS2} Color plots of certain symmetric eigenstates 
$\psi(x_1,x_2)$ 
of the 1D Hamiltonian for $U=1$ and $N=128$ in $x_1$-$x_2$ plane. 
Top panels correspond to two states in the sector $p_+=0$ of total 
momentum with relative level number (of this sector and with states ordered 
by increasing energies) being $l=64$ (left) and $l=44$ (right). Center panels 
correspond 
to $p_+=21\pi/32\approx 2\pi/3$ with $l=64$ (left) and $l=43$ (right).
Bottom panels correspond to $p_+=63\pi/64\approx \pi$ with 
$l=62$ (left) and $l=43$ (right). 
}
\end{figure}

In Fig.~\ref{figS2} we show some examples of 1D eigenstates of 
the quantum Hamiltonian (\ref{eq_quant_Ham}) obtained from (the real 
or imaginary part of) Eq. (\ref{eq_eig_reconstruc}) as explained in 
the last section from symmetrized block eigenstates (with $65$ or $64$ 
levels per symmetrized $p_+$ sector). The parameters are $U=1$, $N=128$ 
and three values of $k=p_+=0$, $p_+\approx 2\pi/3$ and $p_+\approx \pi$.
For states with (near) maximal energy (for their respective $p_+$ sector) 
one sees a 
strong localization on the diagonal indicating a strong localization in the 
relative coordinate $\Delta x$. For relative level numbers close to 
$2/3$ of the maximal level number 
the states extend to the full space except for $p_+\approx\pi$ where 
a small strip close to the diagonal is excluded. This is a clear effect 
of energy miss-match and the very strong interaction on the diagonal 
if compared to the strongly reduced kinetic energy.
The (presence or absence of) change of colors in the center of mass 
direction indicate the periodic oscillations due 
the wave number $p_+$ in this direction. 

\begin{figure}[h]
\begin{center}
\includegraphics[width=0.4\textwidth]{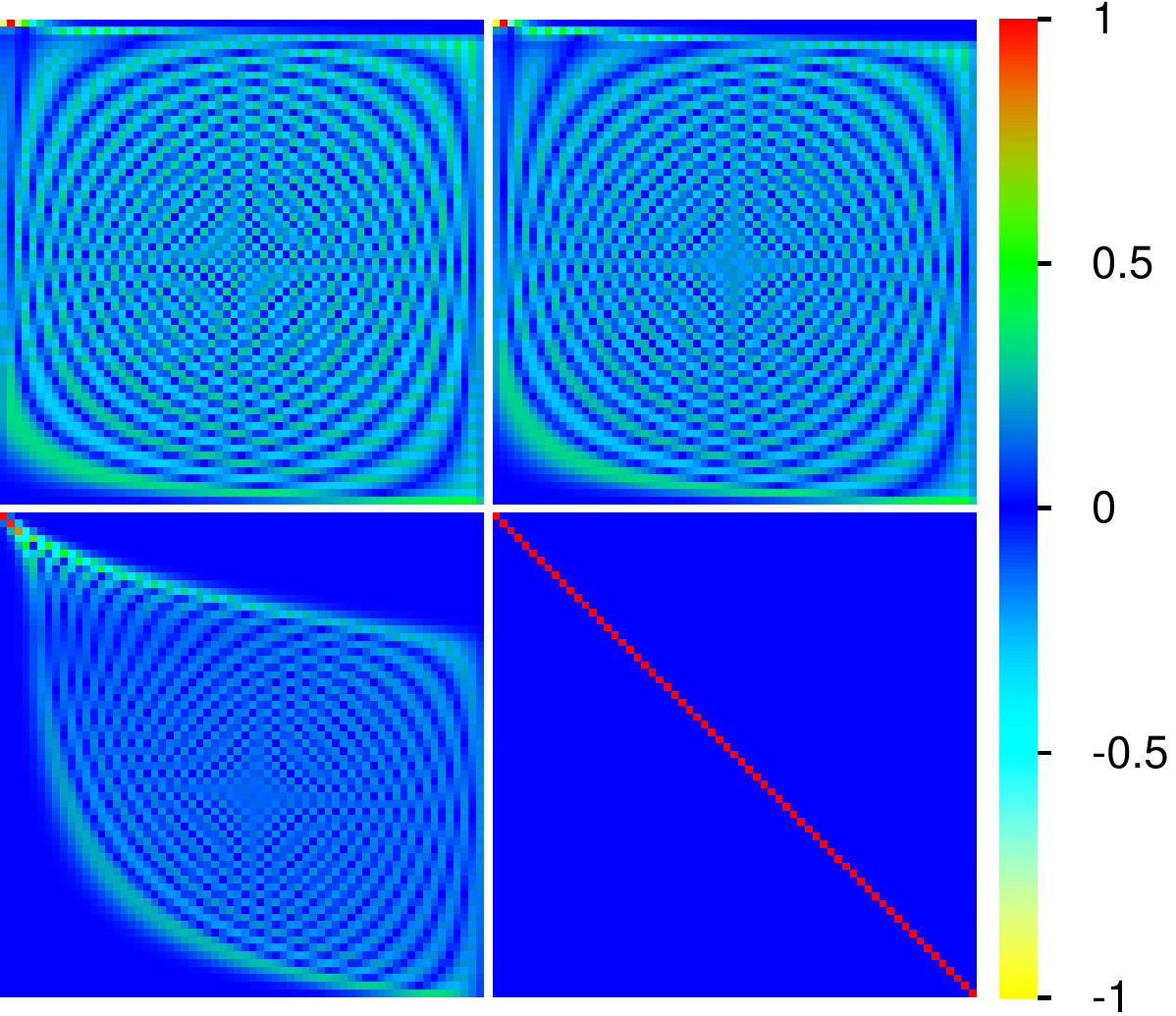}
\end{center}
\caption{\label{figS3} Schematic representation of all symmetric 
block eigenstates for the 1D case and $U=1$, $N=128$ in relative coordinate 
$\Delta x$ corresponding to the horizontal axis and with block level 
number $l$ corresponding to the vertical axis. The different panels 
correspond to different values of the conserved total momentum which are:
$p_+=0$ (top left), $p_+=21\pi/32\approx 2\pi/3$ (top right), 
$p_+=63\pi/64\approx\pi$ (bottom left) and $p_+=\pi$ exactly (bottom right).
}
\end{figure}

Fig.~\ref{figS3} shows a schematic representation of all symmetrized 
1D block eigenstates of certain $p_+$ sectors (for $U=1$, $N=128$). The 
horizontal axis of each panel corresponds to $0\le \Delta x\lesssim N/2$
and the vertical axis to the level number $0\le l\lesssim 64$ 
(bottom/top for lowest/largest sector energies) 
of the shown $p_+$ sectors. 
For $p_+=0$ and $p_+\approx 2\pi/3$ there is at top energies a very 
small number of states with rather strong localization at $\Delta x$ close 
to 0 (values $\Delta x$ close to $N$ are not visible due to the 
symmetrization). Then there is a big majority of states which are completely 
delocalized on the full $\Delta x$ interval with certain wavelength values 
depending on the level number. At lowest energies there is also a very 
small number of states with significant localization close at 
$\Delta x\approx N/2$ which is the minimum of the 1D interaction potential.
For $p_+\approx \pi$ but different from $\pi$ the top (bottom) energy ranges 
of strong localization at $\Delta x\approx 0$ ($\Delta x\approx N/2$) are 
strongly enhanced and the intermediate ``delocalized'' states are quite 
strongly excluded from small $\Delta x$ values due to an energy miss-match 
caused by the very strong relative interaction (if compared to the reduced 
kinetic energy). At $p_+=\pi$ exactly there is perfect localization at 
single $\Delta x$ sites for each state since in this case the kinetic energy 
vanishes exactly and the corresponding block Hamiltonian is already diagonal 
with eigenvalues given by the interaction values $U(\Delta x)$. 

These observations of Fig.~\ref{figS3} explain quite clearly the behavior 
of the time evolution of the wave function in block representation visible 
in the bottom panels of Fig.~\ref{fig1}. For sectors with 
$p_+\approx \pi$ an initial 
block state localized at $\Delta x=1$ will have only contributions from the 
strongly localized block eigenstates at top energies while other extended 
block eigenstates have nearly no amplitude at small $\Delta x$ and do not 
contribute in the initial state (when the latter is expanded in terms 
of block eigenstates). Therefore there is nearly perfect localization for 
these sectors (with $p_+\approx \pi$) also for long times. 
For other values of $p_+$ quite different from $\pi$ the number of localized 
top energy states is reduced. However, they still produce an enhanced 
contribution in the eigenvector expansion of the initial state. Now (most of) 
the other extended block eigenstates are not excluded/forbidden at small 
$\Delta x$ values and they have also a certain contribution in the eigenvector 
expansion, but still with smaller coefficients than the localized 
top energy states. Therefore in the time evolution a quite significant 
fraction of probability stays localized while the complementary fraction 
of density propagates through the whole system. 

\begin{figure}[h]
\begin{center}
\includegraphics[width=0.4\textwidth]{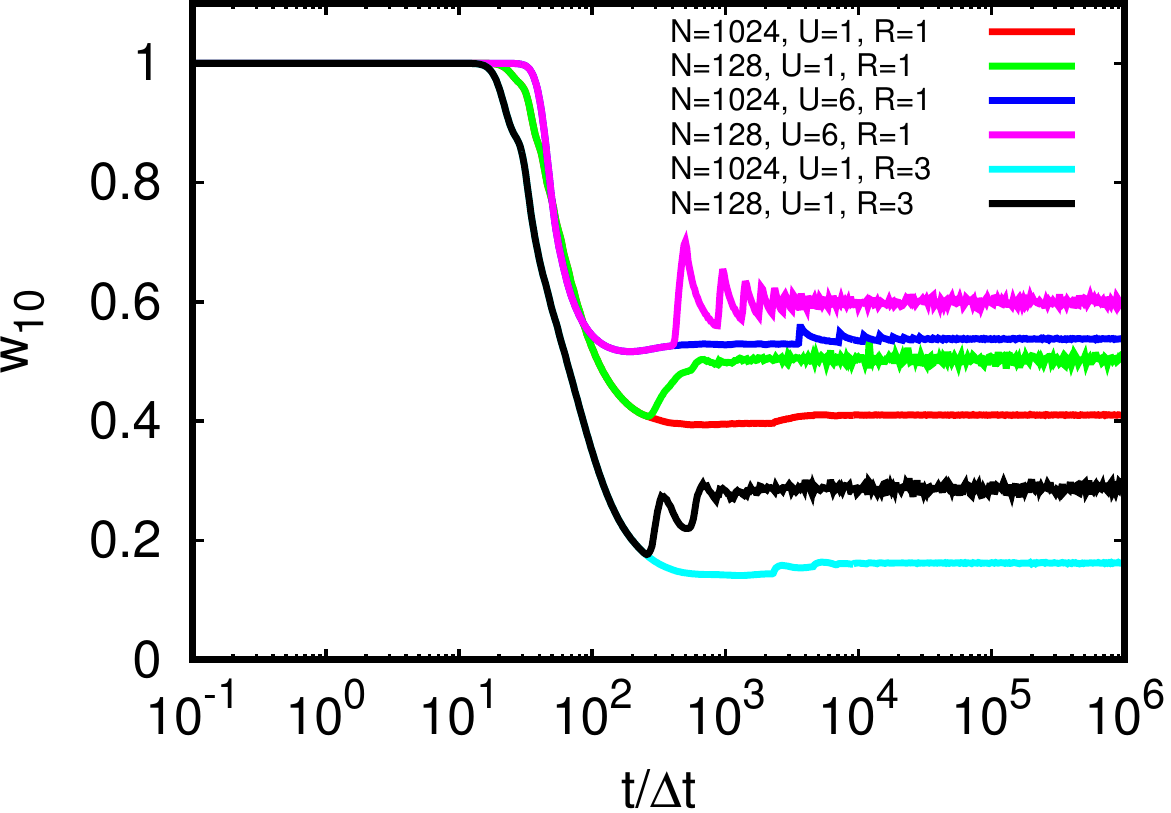}
\end{center}
\caption{\label{figS4} Time dependence of the quantum probability 
$w_{10}$ to find both particles in the relative strip $\Delta\bar x\le 10$ 
for the exact 1D time evolution for different values 
of the parameters $N$, $U$ and $R$ with an initial symmetrized state 
where both particles are localized (at $x_1\approx x_2\approx N/2$)
at distance $\Delta\bar x=R$.
}
\end{figure}

To characterize the quantum probability to form compact electron pairs we 
compute (for the 1D case) the following quantity:

\begin{equation}
\label{eq_w10_1D_def}
w_{10}=\sum_{x_1=0}^{N-1}\sum_{\Delta x=-10}^{10} 
|\psi(x_1,x_1+\Delta x)|^2
\end{equation}
where the sum $x_1+\Delta x$ is taken modulo $N$ and 
$\psi(x_1,x_2)$ represents the wave function (in position representation) 
of the quantum state for which 
we want to compute $w_{10}$, typically a state obtained from the 
exact 1D quantum time evolution. This quantity is just the quantum probability 
to find both particles in the relative strip $\Delta\bar x\le 10$. 
One can also compute $w_{10}$ from the wave function $\bar\psi(k,\Delta x)$ 
in block representation by sums over all $k$ values and $\Delta x$ values such 
that either $\Delta x\le 10$ or $\Delta x\ge N-10$. 

Fig.~\ref{figS4} shows the time dependence of $w_{10}$ using the same 
(type of) time evolution data used for Fig.~\ref{fig1} for different 
parameter combinations with $N=128,1024$, $U=1,6$ for $R=1$ or 
$U=1$ for $R=3$ (here $R$ is the initial distance of the particles). 
We see in all cases that $w_{10}$ saturates at long times at a finite 
value which is significantly above its ergodic values $w_{\rm erg.}=21/N$
(maybe except for $N=128$, $U=1$ and $R=3$) assuming a perfectly uniform 
wave function amplitude. Increasing $R=1$ to $R=3$ clearly reduces 
the pair formation probability while increasing the interaction from 
$U=1$ to $U=6$ increases the value of $w_{10}$. For the data at $N=128$ there 
is also a considerable finite size effect. 

\section{S4. Additional data for the 2D quantum case}

Before presenting some additional data for the 2D quantum case, we provide 
here explicit formulas for the quantities shown in Figs.~\ref{fig4},
\ref{fig5}. 
Assuming that $\psi(x_1,y_1,x_2,y_2)$ is the wave function in position space 
obtained from the Totter formula 2D time evolution top panels of 
Fig.~\ref{fig4} show the density in $\Delta x$-$\Delta y$ plane obtained from:
\begin{equation}
\label{eq_dxdy2Ddens}
\rho_{\rm rel}(\Delta x,\Delta y)=
\sum_{x_1,x_2}|\psi(x_1,y_1,x_1+\Delta x,y_1+\Delta y)|^2
\end{equation}
(with position sums taken modulo $N$) and bottom panels show the density 
in $x_1$-$x_2$ plane obtained from:
\begin{equation}
\label{eq_xx2Ddens}
\rho_{XX}(x_1,x_2)=
\sum_{y_1,y_2}|\psi(x_1,y_1,x_2,y_2)|^2\ .
\end{equation}
The quantum probability $w_{\Delta R}$ to find both 
electrons at $\Delta\bar x\le \Delta R$ and $\Delta\bar y\le \Delta R$ is 
computed from:
\begin{equation}
\label{eq_wDeltar}
w_{\Delta R}=\sum_{\Delta\bar x\le \Delta R,\Delta\bar y\le \Delta R}
\rho_{\rm rel}(\Delta x,\Delta y)
\end{equation}
(with $\Delta\bar x=\min(\Delta x,N-\Delta x)$ and similarly for $\Delta y$). 
For the case of exact 2D block time evolution $w_{\Delta R}$ is computed as 
in \eqref{eq_wDeltar} but with $\rho_{\rm rel}(\Delta x,\Delta y)$ 
replaced by $|\bar\psi(p_{+x},p_{+y},\Delta x,\Delta y)|^2$ (at certain 
given values of $p_{+x},p_{+y}$). 
Here we mostly use the quantity $w_{10}$ corresponding to 
$\Delta R=10$ but below we also show some data for $w_2$ 
corresponding to $\Delta R=2$. 

\begin{figure}[h]
\begin{center}
\includegraphics[width=0.4\textwidth]{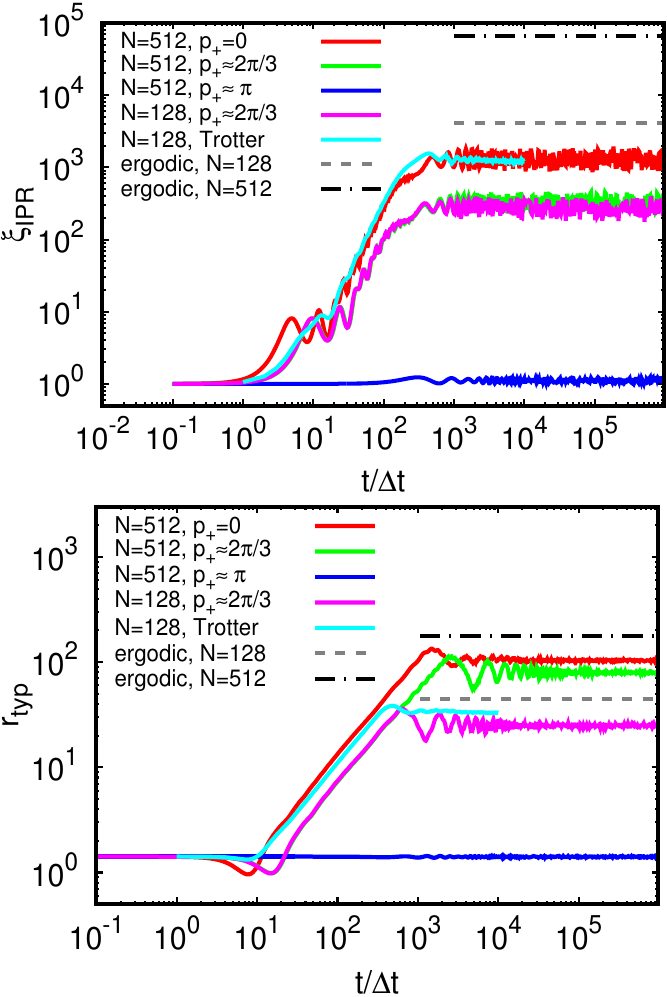}
\end{center}
\caption{\label{figS5} Time dependence of the inverse participation ratio 
$\xi_{\rm IPR}$ (see Eq. \eqref{eq_iprdef}; top panel) and 
typical distance $r_{\rm typ}$ (see Eq. \eqref{eq_rtypdef}; 
bottom panel) computed from the density in $\Delta x$-$\Delta y$ plane for 
the same cases and raw data 
of Fig.~\ref{fig4}. The dashed grey (dashed-dotted black) line shows the 
ergodic value for $N=128$ ($N=512$) assuming a uniform density 
in $\Delta x$-$\Delta y$ plane given by $\xi_{\rm IPR,erg}=(N/2)^2$ 
and $r_{\rm typ,erg}=0.692\times (N/2)$.
}
\end{figure}

In Fig.~\ref{figS5}, we show for the same cases and raw data (same 
2D time evolution wave functions) used for Fig.~\ref{fig4} 
the time dependence of the inverse 
participation ratio defined by:
\begin{equation}
\label{eq_iprdef}
\xi_{\rm IPR}=\left(
\sum_{\Delta x,\Delta y}\rho_{\rm rel,sym}^2(\Delta x,\Delta y)
\right)^{-1}
\end{equation}
($\rho_{\rm rel,sym}$ is the symmetrized density with respect to the 
two symmetries $\Delta x\leftrightarrow N-\Delta x$ and 
$\Delta y\leftrightarrow N-\Delta y$ obtained from $\rho_{\rm rel}$)
and the typical particle distance:
\begin{equation}
\label{eq_rtypdef}
r_{\rm typ}
=\exp[\langle\ln(\sqrt{\Delta \bar x^2+\Delta \bar y^2}+1)\rangle]-1
\end{equation}
where the average $\langle\cdots\rangle$ is done with respect to 
the symmetrized density $\rho_{\rm rel,sym}$. 

The inverse participation ratio provides roughly the number of sites 
over which $\rho_{\rm rel,sym}(\Delta x,\Delta y)$ is mostly concentrated or 
localized. The saturation values at long times well below the ergodic value 
clearly indicate a considerable pair formation probability confirming the 
findings of Fig.~\ref{fig5}. For $p_+\approx \pi$ (case of near perfect 
localization) one can see values slightly above unity 
($\xi_{\rm IPR}\lesssim 1.2$) for all time scales. Note that for this 
case $w_{10}$ shown in Fig.~\ref{fig5} takes the value 1 with 
numerical precision (error below $10^{-14}$). 

The other quantity $r_{\rm typ}$ is mostly dominated by the fraction 
of density associated to free electron propagation. Therefore the ratio 
$r_{\rm typ}/r_{\rm typ,erg}$ corresponds to the complementary 
non-pair formation probability and indeed the quantity 
$1- r_{\rm typ}/r_{\rm typ,erg}$ behaves qualitatively similarly as $w_{10}$. 
For the case of near perfect localization at $p_+\approx \pi$ we have for all 
time scales $r_{\rm typ}\approx \sqrt2$ (corresponding to 
$\Delta x=\Delta y=1$) with some slight fluctuations 
$\pm 0.02$ at long times.

Globally, we believe that $w_{10}$ is more useful and suitable than 
$\xi_{\rm IPR}$ or $r_{\rm typ}$ to describe the pair formation probability 
which is the reason why we have chosen to show in the main part of this work 
the time dependence of $w_{\rm 10}$ in Fig.~\ref{fig4}. 

\begin{figure}[h]
\begin{center}
\includegraphics[width=0.4\textwidth]{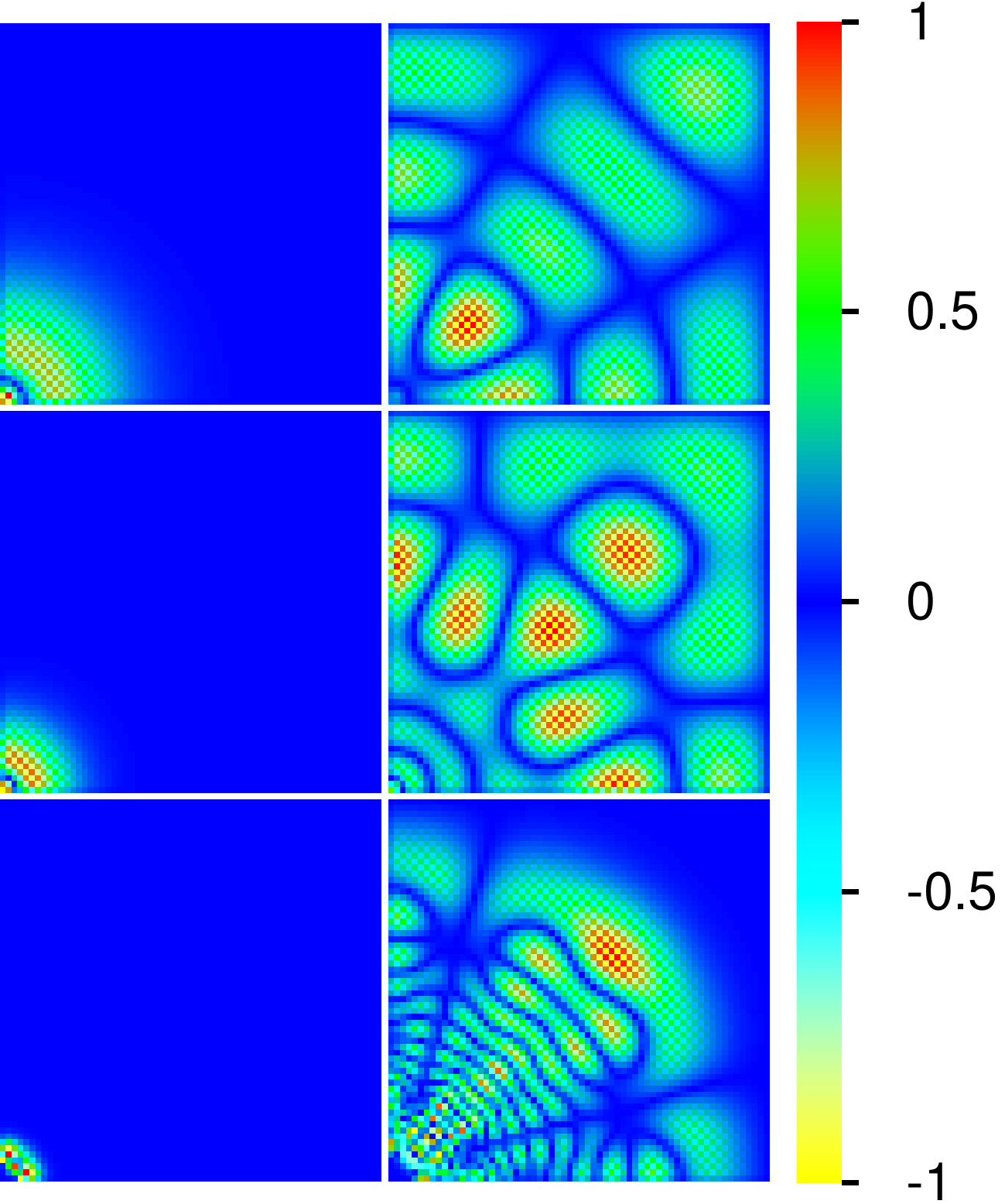}
\end{center}
\caption{\label{figS6} 
Certain 2D (totally symmetrized) block eigenstates in 
$\Delta x$-$\Delta y$ plane for $U=2$, $N=128$. 
Top panels correspond to conserved total momentum $p_+=p_{+x}=p_{+y}=0$ and 
block level numbers $l=2143$ (left) and $l=2135$ (right).
Center panels correspond to $p_+=21\pi/32\approx 2\pi/3$, 
$l=2143$ (left), $l=2131$ (right). Bottom panels correspond to 
$p_+=63\pi/64\approx\pi$, $l=2072$ (left), $l=1991$ (right).
}
\end{figure}

Some examples of localized and delocalized 2D block eigenfunctions for the 
three sectors $p_+=p_{+x}=p_{+y}=0$, $p_+\approx 2\pi/3$ and 
$p_+\approx\pi$ and $N=128$, $U=2$ are presented in Fig.~\ref{figS6}. As in 
1D top energy eigenstates (with respect to their $p_+$ sector) are localized 
at $\Delta x\approx \Delta y\approx 0$ and lower energy eigenstates are 
delocalized. For $p_+=0$ and $p_+\approx 2\pi/3$ the number of localized 
states is very small and already at sector level numbers slightly below 
their maximal value delocalization sets in. For $p_+\approx\pi$ the number 
of localized states is larger and for the delocalized states the zone 
close to $\Delta x\approx \Delta y\approx 0$ is forbidden such that these 
states do not contribute in the time evolution if the initial state is 
localized at $\Delta x=\Delta y=R$ for some small value of $R=1$ or $R=3$.
These findings are very similar to the 1D case (see above discussion 
of Figs.~\ref{figS3},\ref{figS4}) and help to understand the 2D time 
evolution results shown in Figs.~\ref{fig3}-\ref{fig5}.

One can also see, especially for the delocalized eigenstates,
a spatial node structure 
indicating a rough and very approximate angular momentum conservation. 
However, this is not exact due to the square form of available space, discrete 
lattice structure and also the cosinus 2D band form of kinetic energy (for 
the relative coordinates). Furthermore, our limitation to totally symmetrized 
block eigenstates (with respect to the discrete symmetries explained in 
Sec. S2) plays a role here.

\begin{figure}[h]
\begin{center}
\includegraphics[width=0.4\textwidth]{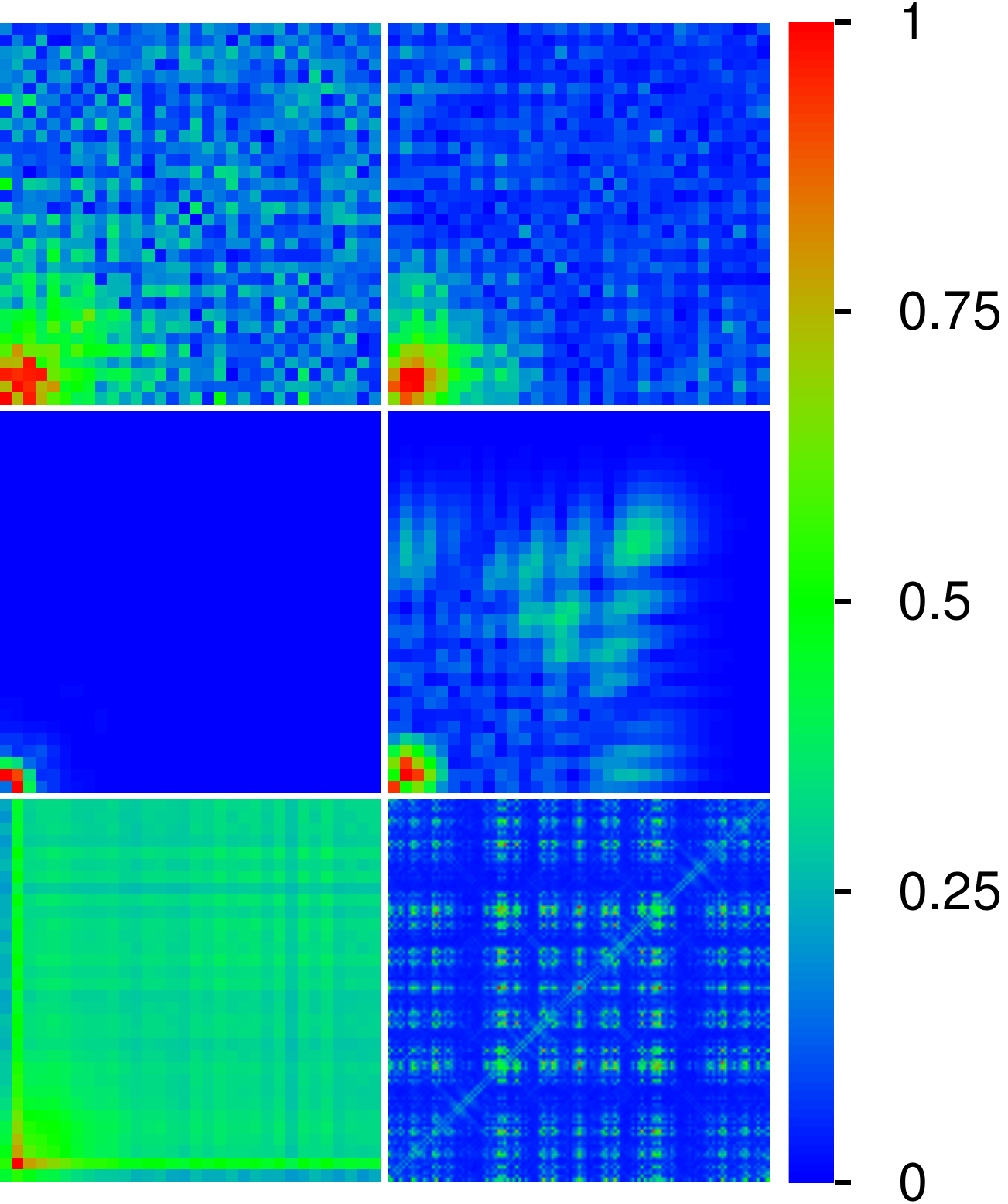}
\end{center}
\caption{\label{figS7} Certain wave function densities for different 
cases of the 2D time evolution and weak interaction $U=0.5$. 
Top and center panels correspond to the same type of quantity as 
shown in Fig.~\ref{fig3} with $p_+=p_{+x}=p_{+y}=0$ (top left), 
$p_+=85\pi/128\approx 2\pi/3$ (top right), $p_+=255\pi/256\approx \pi$ 
(center left), all three for $N=512$ and $t=10^5\,\Delta t$, 
and $p_+=63\pi/64\approx \pi$, $N=128$, $t=3090\,\Delta t$ (center right).
Bottom panels correspond to the Trotter formula approximation for 
$N=128$ and $t=9487\,\Delta t$ with left panel showing 
the density in relative coordinates obtained from 
a sum over $x_1$ and $y_1$ and with right panel showing 
the density in $x_1$-$x_2$ plane obtained from a sum over $y_1$ and $y_2$. 
All panels (except bottom right) show to a zoomed region 
$0\le \Delta x,\Delta y<32$ in $\Delta x$-$\Delta y$ plane. 
At \cite{ourwebpage} videos corresponding to the cases of all panels of this 
figure (and also for $N=128$ with $p_+=0$ and $p_+=21\pi/32\approx 2\pi/3$) 
are provided. 
}
\end{figure}

The 2D pair formation effect still exists at the smaller interaction 
value $U=0.5$ even though it is less pronounced as can be seen in 
Fig.~\ref{figS7} showing certain wave function densities obtained from 
both types of 2D time evolution methods. For $N=128$ the Trotter formula 
results indicate the formation of a very slight diagonal in the density 
$\rho_{XX}(x_1,x_2)$ and a modest localization effect in 
$\rho_{\rm rel}(\Delta x,\Delta y)$. However, here the density weight 
corresponding to free electron propagation is quite high.
For $N=128$ this is also confirmed by the 2D block time evolution results 
(not shown in Fig.~\ref{figS7} except for one special case) but 
for $N=512$ the density related to free propagation is quite strongly 
reduced in comparison to $N=128$. 
As can be seen in both top and left center panels of Fig.~\ref{figS7} 
for $N=512$ there is a considerable density for the compact electron pair. 
However, for $p_+=0$ ($p_+\approx 2\pi/3$) one still clearly see a quite 
strong (quite modest) density related to free propagation while 
for $p_+\approx \pi$ the latter is absent. 
The right center panel shows the case $N=128$ and $p_+\approx \pi$ 
(strong localization case at $N=512$) at a certain intermediate time when one 
can briefly see a density related to free propagation which is however smeared 
out at longer times.

\begin{figure}[h]
\begin{center}
\includegraphics[width=0.4\textwidth]{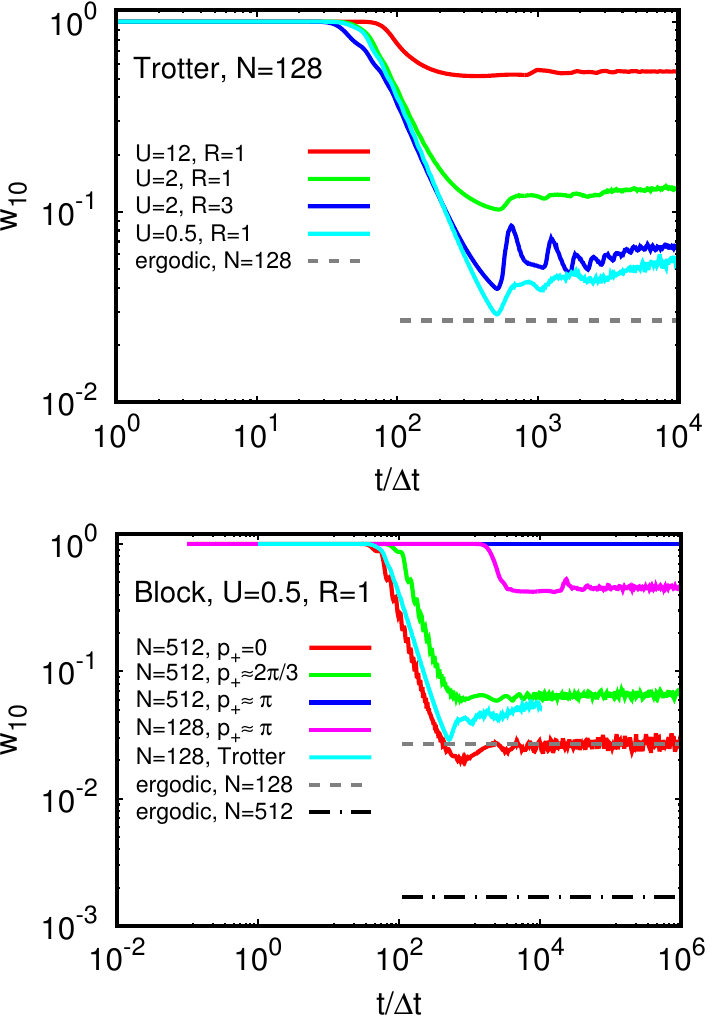}
\end{center}
\caption{\label{figS8} Time dependence of the quantum probability 
$w_{10}$ to find both particles in the relative square $\Delta \bar x\le 10$, 
$\Delta \bar y\le 10$. Top panel corresponds to the full space 
2D time evolution using the Trotter formula approximation for $N=128$ 
and different values of $U$ and initial particle distance $R$.
Bottom panel corresponds to $U=0.5$, $R=1$ and 
compares certain cases of the exact 2D block time evolution 
in certain sectors of conserved momenta $p_+=p_{+x}=p_{+y}$ with the result 
obtained from the Trotter formula approximation. The precise 
values of $p_+$ are~: $p_+=0$, $p_+=85\pi/128\approx 2\pi/3$, 
$p_+=255\pi/256\approx\pi$ all three for $N=512$ and 
$p_+=63\pi/64\approx \pi$ for $N=128$. 
The top blue line has at $t=10^6$ a numerical 
of $0.99918$. 
In both panels the grey dashed horizontal line corresponds to the value of 
$w_{\rm erg.}=(21/128)^2\approx 0.0269$ assuming a constant (ergodic) 
density on a square of size $N=128$. 
In bottom panel the black dashed-dotted horizontal line 
corresponds to the value of $w_{\rm erg.}=(21/512)^2\approx 0.00168$ 
assuming a constant density on a square of size $N=512$. 
}
\end{figure}

The qualitative findings of Fig.~\ref{figS7} for $U=0.5$ are confirmed by 
the time dependence of $w_{10}$ shown in the lower panel of Fig.~\ref{figS8} 
for the cases of Fig.~\ref{figS7}. For $N=512$ and $p_+=0$ or 
$p_+\approx 2\pi/3$ the saturation values of $w_{10}$ 
are below 0.1 but still significantly above the ergodic value. 
For $p_+\approx\pi$ and $N=512$ the saturation value $w_{10}=0.99988$ 
is nearly maximal while for $N=128$ it is lower but still quite elevated, 
$w_{10}\approx 0.45$, showing a certain finite size effect for this case. 
The saturation of the Trotter formula case at $N=128$ is $w_{10}\approx 0.05$ 
which is rather low but still roughly twice the ergodic value which is 
$0.027$. 

The top panel of Fig.~\ref{figS7} shows the time dependence of $w_{10}$ for 
the Trotter formula case and other parameter values $U=0.5,\,2,\,12$ at $R=1$
and $U=2$ at $R=3$ with $R$ being the initial value of 
$\Delta\bar x=\Delta\bar y$. As 
in 1D (see Fig.~\ref{figS4}) stronger (lower) interaction increases 
(reduces) the pair formation probability and when $R$ is increased the 
pair formation probability is reduced. 

\begin{figure}[h]
\medskip

\begin{center}
\includegraphics[width=0.4\textwidth]{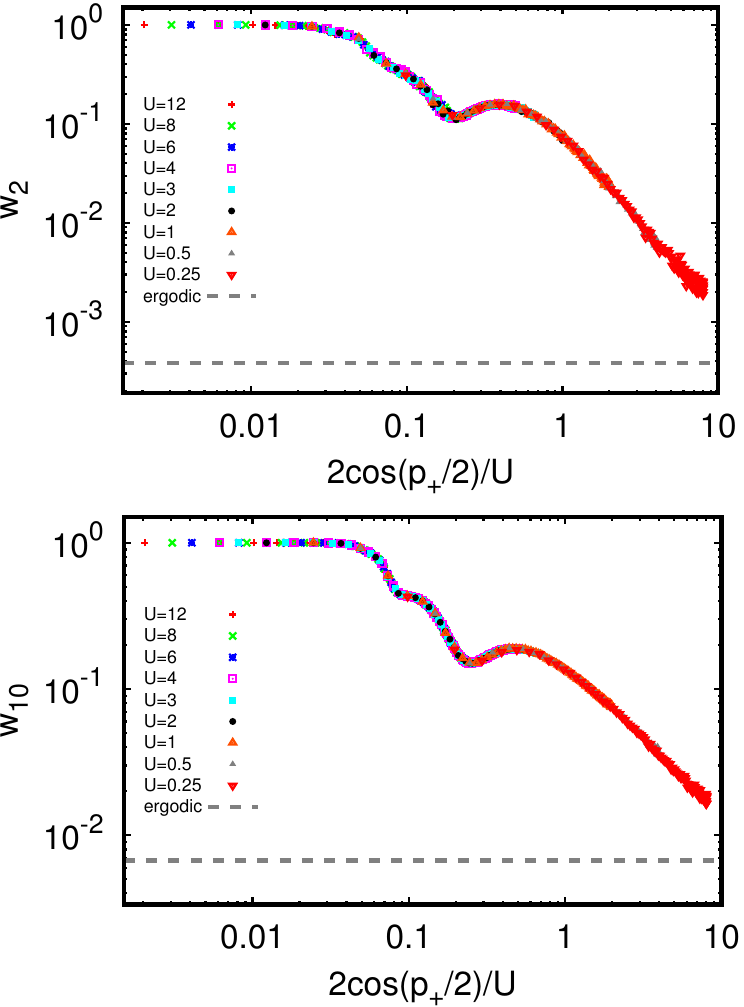}
\end{center}
\caption{\label{figS9} 
Dependence of $w_{\Delta R}$ with $\Delta R=2$ ($\Delta R=10$) 
in top (bottom) panel to find both particles in the relative square 
$\Delta \bar x\le \Delta R$ on the scaling parameter $2\cos(p_+/2)/U$ 
for different interaction values $0.25\le U\le 12$, $N=256$ 
and $p_+=p_{+x}=p_{+y}\in[0,\pi]$. 
The dashed gray line corresponds to the ergodic probability 
$w_{\rm erg.}=((2\Delta R+1)/256)^2$ assuming a constant 
density on a square of size $N=256$.
The value of $w_{\Delta R}$ has been computed from the exact 2D time evolution 
in different $p_+$ sectors using a time average over 21 time values $t$ 
with logarithmic density such $10^4\le t/\Delta t\le 10^6$.
}
\end{figure}

\begin{figure}[h]
\begin{center}
\includegraphics[width=0.4\textwidth]{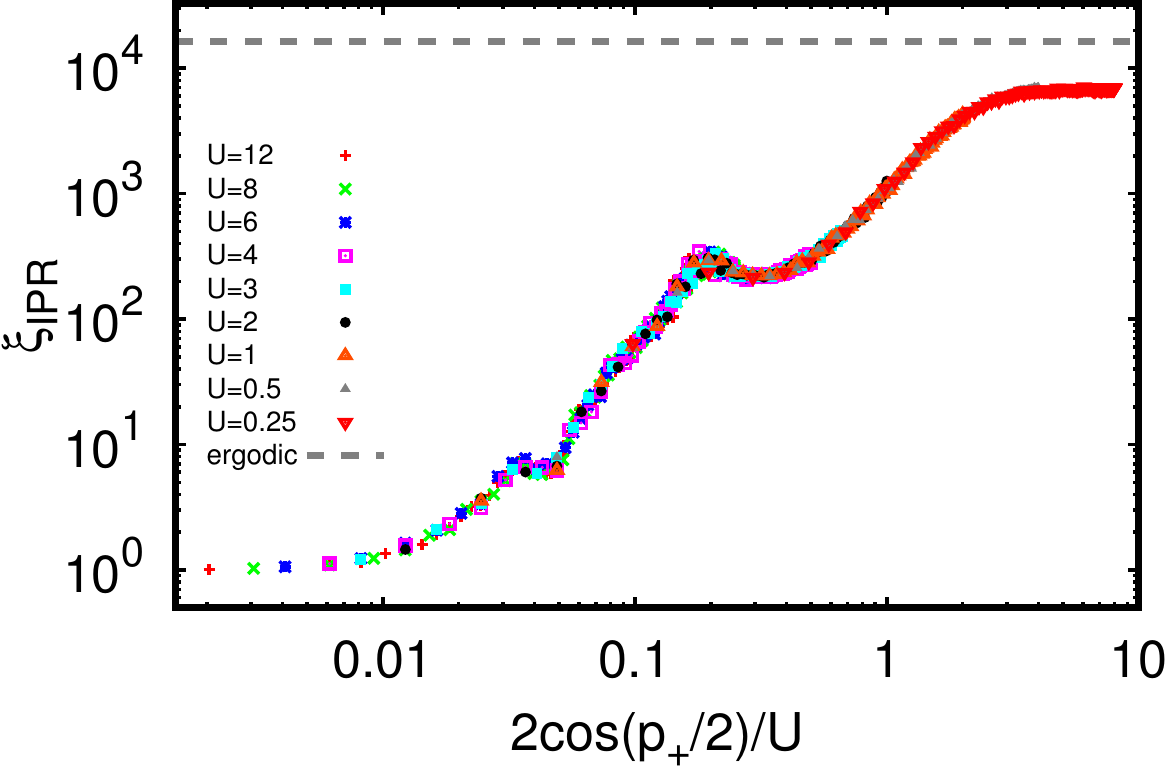}
\end{center}
\caption{\label{figS10} Dependence of the 2D inverse participation ratio 
$\xi_{\rm IPR}$ in symmetrized $\Delta x$-$\Delta y$ plane 
on the scaling parameter $2\cos(p_+/2)/U$. 
The dashed grey line corresponds to the ergodic value 
$\xi_{\rm IPR,erg.}=(N/2)^2\approx 1.6\times 10^4$ assuming a uniform 
distribution on a square of size $N/2=128$ (due to symmetrization). 
$\xi_{\rm IPR}$ has been computed from the same raw data (time dependent 
wave functions), using the same parameters and same time average as in 
Fig.~\ref{figS9}. 
}
\end{figure}

In Fig.~\ref{figS9} we show for $N=256$ the dependence of the saturation value 
of $w_{\Delta R}$ for $\Delta R=2$ (top panel) and $\Delta R=10$ 
(bottom panel), computed as the average 
over 21 time values $t$ with $10^4\le t/\Delta t\le 10^6$, on the 
scaling parameter $C=2\cos(p_+/2)/U$ (here we only consider $0\le p_+\le \pi$ 
such that $C\ge 0$). This parameter 
is the ratio of the hopping matrix 
element and the interaction amplitude of the effective 2D {\em one-particle} 
tight binding model of the block at $p_+=p_{+x}=p_{+y}$ (see Sec.~S2 
for details). 
We have also computed the same scaling curves for $N=512$ with a reduced 
density of data points. These curves are identical to (slightly below) the 
case $N=256$ for $C\lesssim 2$ ($C\gtrsim 2$) with a rough factor of 
0.5 at $C=8$.

The scaling dependence is quite obvious and clearly confirms the role 
of the total conserved momentum $p_+$ for the kinetic energy amplitude of 
the relative coordinates. At small values of $C\ll 1$ we 
have perfect localization (perfect pair formation) with 
$w_2\approx w_{10}\approx 1$ and then the 
curves decrease down to a modest local minimum at $C\approx 0.2$ which is 
followed by a local maximum at $C\approx 0.3$. 
Then they continue to decrease down to the value of $2.2\times 10^{-3}$ 
($1.8\times 10^{-2}$) for $w_2$ ($w_{10}$) at $C=8$. These minimal 
values are still clearly above the ergodic values $3.8\times 10^{-4}$ 
($6.7\times 10^{-3}$). 

Fig.~\ref{figS10} shows the same type of scaling curve but for the 
inverse participation ratio $\xi_{\rm IPR}$ which has a value of $\approx 1$ 
for $C\ll 1$ for the case of perfect localization (perfect pair formation). 
Then the curve increases up to a modest local maximum at $C\approx 0.2$ 
followed by a local minimum at $C\approx 0.3$. 
After this it continues to increase until it saturates for $C\gtrsim 3$ 
at the value $\xi_{\rm IPR}\approx 6.9\times 10^3$ which is clearly 
below the ergodic value $1.6\times 10^4$. 

\section{S5. Videos of TIP time evolution  in 2D}

Two videos of the 2D time evolution for the cases of (i) Fig.~\ref{fig3} (videoforfig3.avi), 
top right panel (exact 2D block time evolution of sector 
$p_+=21\pi/32\approx 2\pi/3$) and (ii) Fig.~\ref{fig4} (videoforfig4.avi), bottom panels 
(density in $x_1$-$x_2$ plane obtained from the Trotter formula 2D 
time evolution) are provided. 
For (i) the video is composed of 702 images (25 images per second of video) 
at time values $t_0=0$ and 
$t_i=\Delta t\times 10^{(i-101)/100}$ for $1\le i\le 701$ (with 
$\Delta t=1/B_2=1/(16+U)$) corresponding to a logarithmic time scale 
in the range $\Delta t\times 10^{-1}\le t\le \Delta t\times 10^6$. 
For (ii) the video is composed of 464 images at integer multiples of 
$t_i=l_i\Delta t$ with $l_i=i$ for $0\le i\le 87$ and 
roughly $l_{i+1}\approx 10^{1/100}\,l_i$ for $87\le i< 463$ (such 
that $l_{463}=10^4$) 
corresponding roughly to a logarithmic (linear) time scale 
for $t/\Delta t>87$ ($t/\Delta t\le 87$) in the range 
$\Delta t\le t\le \Delta t\times 10^4$. 

At our web page \cite{ourwebpage} 
\url{http://www.quantware.ups-tlse.fr/QWLIB/coulombelectronpairs} 
further videos corresponding to the panels 
of Fig.~\ref{fig1} (exact full space 1D time evolution), Fig.~\ref{figS7} 
(both cases of 2D time evolution for $U=0.5$), 
and other panels of Figs.~\ref{fig3},\ref{fig4} are available. 
The time scales are 
always as in the cases (i) (exact 1D, with modified $\Delta t=1/B_1=1/(8+U)$, 
or block 2D time evolution) or (ii) (Trotter formula 2D time evolution). 

\end{document}